\documentclass[aps,prb,reprint,superscriptaddress,amsmath,amssymb,showpacs]{revtex4-1}
\usepackage{graphicx}
\usepackage{bbm}

\date{\today}                  

\begin{document}

\title{Transient dynamics of open quantum systems}

\author{Oleksiy Kashuba}
\email[Email: ]{kashuba@physik.rwth-aachen.de}
\affiliation{Institut f\"ur Theorie der Statistischen Physik, RWTH Aachen, 
52056 Aachen, Germany and JARA - Fundamentals of Future Information Technology}
\author{Herbert Schoeller}
\affiliation{Institut f\"ur Theorie der Statistischen Physik, RWTH Aachen, 
52056 Aachen, Germany and JARA - Fundamentals of Future Information Technology}

\begin{abstract}
We present a renormalization group (RG) method which allows for
an analytical study of the transient dynamics of open quantum systems
on all time scales. Whereas oscillation frequencies and decay rates of exponential time 
evolution follow from the fixed point positions, the long-time 
behavior of pre-exponential functions is related to the scaling behavior around
the fixed points. We show that certain terms of the RG flow are only cut off by inverse time,
which leads to a difference between infrared and ultraviolet scaling. An 
evaluation for the Ohmic spin boson model at weak damping reveals significant deviations from 
previous predictions in the long-time regime. We propose that 
weak-coupling problems for stationary quantities can in principle turn into strong-coupling 
ones for the determination of the long-time behavior.
\end{abstract}

\pacs{05.10.Cc, 05.30.-d, 05.30.Jp, 73.23.-b}

\maketitle
The time dynamics of a small strongly interacting quantum system coupled to 
noninteracting large reservoirs is a fundamental issue in nonequilibrium
statistical mechanics. The prototype is a two-level system coupled to an
environment,\cite{Leggett_87,Weiss_12} which is of particular interest in
quantum information processing.\cite{Peskill_98} A typical setup is the
one of transient dynamics: The system and environment are decoupled for 
times $t<0$ and the coupling is switched on suddenly at $t=0$. The time
evolution of the reduced density matrix $\rho_t$ of the local system will then 
be characterized for $t>0$ by a series of terms, each of which will generically
be of the form of an exponential together with a pre-exponential function
$\rho_t=\sum_n F^n_t\,\exp{(-iz_nt)}$. Here, $z_n=\Omega_n-i\Gamma_n$ 
consists of an oscillation frequency $\Omega_n$ and a decay rate $\Gamma_n$,
where one of the scales $z_n$ will be zero characterizing the stationary state
$\rho_{\text{st}}$ for $t\rightarrow\infty$. Besides the calculation of $\rho_{\text{st}}$ 
and $z_n$, the main challenge lies in the analysis of the pre-exponential functions $F^n_t$ 
on all time scales. Although interesting field-theoretical\cite{exact} and
numerical techniques\cite{numerics} have been developed to study the time dynamics,
the precise form of pre-exponential functions has not been addressed so far. 
Promising tools for this purpose are  
perturbative renormalization group (RG) methods for nonequilibrium problems, like
the flow equation method,\cite{flow_equation,Hackl_etal_09} real-time RG (RTRG)\cite{Schoeller_09,Pletyukhov_Schuricht_Schoeller_10,IRLM_10_11,Pletyukhov_Schoeller_12}
and functional RG\cite{frg} techniques, or combinations of the latter two.\cite{Kennes_etal_13}
The RTRG method allows for an analytical study on all time scales, provided that the RG flow stays in
the weak-coupling regime. The time dynamics is related to the
density matrix $\rho(E)$ in Laplace space, where the exponential scales $z_n$ 
are the singularities of $\rho(E)$ in the complex plane and the pre-exponential functions
can be determined from branch cut integrals starting at these singularities. 
In Ref.~\onlinecite{Pletyukhov_Schoeller_12}
a RG approach has been proposed by using the Laplace variable $E$ itself as flow
parameter (called E-RTRG in the following), where the singularities $z_n$ are given by 
the fixed points of the RG flow and the long-time behavior of pre-exponential functions 
can be related to the scaling behavior around the fixed points. 

In this Rapid Communication we will combine E-RTRG with a new parametrization of
the effective Liouvillian in terms of slowly varying logarithmic functions and provide
a discussion of the generic time evolution on all time scales. The main result is the insight 
that, for the determination of pre-exponential functions, certain terms of the RG flow 
are only cut off by the energy scale of inverse time $1/t$. This is in contrast to stationary 
quantities, where it has been proposed\cite{Rosch_etal_01_03,Glazman_Pustilnik_05} and 
microscopically shown\cite{Kehrein_05,Schoeller_Reininghaus_09} that {\it all} terms of the RG flow 
are cut off by decay rates. As a consequence, we find that the long-time
behavior is generically quite different from that discussed in 
Refs.~\onlinecite{Pletyukhov_Schuricht_Schoeller_10,IRLM_10_11} at intermediate and short times.
To show this explicitly we will apply our method to 
the Ohmic spin boson model at weak damping, which turns out to be a weak-coupling problem
even close to the fixed points.
For the diagonal components of the density matrix, we find that the power-law
exponent for the scaling behavior of the pre-exponential function agrees with perturbation theory,
in contrast with that predicted by the noninteracting blip approximation (NIBA).\cite{Leggett_87,Weiss_12}
For the nondiagonal elements we find a rather complex scaling behavior which differs from that of 
perturbation theory. We expect similar deviations to occur for other models of open quantum systems as well.
In particular, for certain problems, e.g., 
the antiferromagnetic nonequilibrium Kondo model at large bias voltage, it may even turn out that
the renormalized vertices are small for the calculation of stationary quantities, but 
become large close to the fixed points $z_n$, i.e., a weak-coupling problem for stationary
quantities can turn into a strong-coupling one for the study of the long-time behavior.
This opens up another class of interesting problems for the future. 

{\it Generic discussion.}--- For a discrete quantum system coupled to noninteracting reservoirs at time $t=0$,
it can be shown\cite{Liouvillian_general} that the time evolution of the reduced density matrix 
$\rho_t$ of the local system follows from $i\dot{\rho}_t=\int_0^t dt' L_{t-t'}\rho_{t'}$,
where $L_t$ is an effective Liouvillian superoperator acting on local operators. 
Defining the Laplace transform via $A(E)=\int_0^\infty dt e^{iEt}A_t$, one obtains the formal solution 
$\rho(E)=i\Pi(E)\rho_{t=0}$ with the propagator $\Pi(E)={1\over E-L(E)}$ and the time evolution
follows from the inverse Laplace transform
\begin{equation}
\label{eq:pi}
\rho_t\,=\,{i\over 2\pi}\int_{-\infty+i0^+}^{\infty+i0^+} dE\,e^{-iEt}\,\Pi(E)\,\rho_{t=0}\quad.
\end{equation}
Reference~\onlinecite{Schoeller_09} shows how to calculate the Liouvillian $L(E)$ from a 
diagrammatic expansion in some appropriately defined dimensionless system-bath coupling $\alpha$.
The propagator $\Pi(E)$ is an analytic function in the upper half of the complex plane with poles 
at $z_k^p$ ($k=\text{st},0,\pm 1,\pm 2,\dots$) in the lower half of the complex plane. $k=\text{st}$
denotes the zero pole $z_{\text{st}}^p=0$, which defines the stationary state.
At zero temperature,\cite{comment_a}
additional nonanalytic features arise from branch cuts starting at the singularities
$z_n=z_k^p + \Delta\mu_n$, which are generically given by the pole positions 
shifted by some linear combination $\Delta\mu_n$ of the chemical potentials of the reservoirs.
They arise since the Liouvillian depends logarithmically on terms $\sim\alpha\ln({D\over E-z_n})$, 
generated by ultraviolet divergencies in the band width $D$ of the reservoirs. 
The logarithmic divergencies can be systematically resummed by using E-RTRG.\cite{Pletyukhov_Schoeller_12}
The RG equations express derivatives of the Liouvillian $L(E)$ by a diagrammatic series in terms of
effective vertices, which is free of divergences and can be systematically truncated for weak-coupling
problems. Solving the RG equations along the path $E=z_n+i\Lambda\pm 0^+$ in the complex plane  
starting at $\Lambda=D$, one can tune the positions of the branch cuts to 
$E=z_n-ix$, $x>0$.

For a discussion of the generic time evolution, it is very helpful to use the form
\begin{equation}
\label{eq:L_decomposition}
L(E)\,=\,L_\Delta(E)\,+\,E\,L'(E)\quad,
\end{equation}
where $L_\Delta(E)$ and $L'(E)$ are slowly varying logarithmic functions. This form
is valid in the universal regime $|E|\ll D$. The nonuniversal regime $|E|\gtrsim D$, which 
corresponds to the {\it ultrashort time regime} $t\lesssim 1/D$, is not of interest here since it depends
on the microscopic details of the high-energy cutoff function. 
In the supplementary material\cite{SM} it is shown how the E-RTRG method can be used to 
obtain RG equations for $\tilde{L}_\Delta(E)=Z'(E)L_\Delta(E)$ and the
$Z$-factor superoperator $Z'(E)=1/[1-L'(E)]$. With these quantities one can express
the time evolution (\ref{eq:pi}) in a form where the slowly varying logarithmic parts are
explicitly shown
\begin{eqnarray}
\label{eq:pi_tilde}
\rho_t&=&{i\over 2\pi}\int_{-\infty+i0^+}^{\infty+i0^+} dE\,e^{-iEt}\,\tilde{\Pi}(E)Z'(E)\,\rho_{t=0}\\
\label{eq:pi_tilde_modes}
&=&\sum_k {i\over 2\pi} \int_\gamma dE\,
{e^{-iEt}\over E-\lambda_k(E)}P_k(E)Z'(E)\,\rho_{t=0}\,,
\end{eqnarray}
where $\tilde{\Pi}(E)={1\over E-\tilde{L}_\Delta(E)}$. $\lambda_k(E)$ are the  
eigenvalues of $\tilde{L}_\Delta(E)$ with projectors $P_k(E)$. The poles $z_k^p$ follow
from $z_k^p=\lambda_k(z_k^p)$. The integral is performed by closing the integration contour $\gamma$
in the lower half of the complex plane.  
Using the general expressions (\ref{eq:pi}), (\ref{eq:pi_tilde}) and (\ref{eq:pi_tilde_modes}), 
the typical time dynamics can be obtained as follows:

For {\it short times}, $t\ll 1/|z_n|$,
one can replace $E\rightarrow 1/t$ in the logarithmic functions $\tilde{L}_\Delta$ and $Z'$. This gives
for Eq.~\eqref{eq:pi_tilde} the result $\rho_t\approx e^{-i\tilde{L}_\Delta(1/t)t}Z'(1/t)\rho_{t=0}$. Expanding the 
exponential, one finds in leading order that the scaling behavior at small times follows from the scaling
behavior of $Z'(1/t)$ at large energies. This is the poor man scaling regime, where the cutoff scales
$z_n$ are unimportant. As a result, one obtains universal short-time behavior, which has been reported, 
e.g., for spin boson\cite{Leggett_87,Weiss_12} and Kondo models 
\cite{Hackl_etal_09,Pletyukhov_Schuricht_Schoeller_10}.

For {\it intermediate and long times}, $t\gtrsim 1/|z_n|$, we consider each integral separately around the branch cut
at $E=z_n-ix$, $x>0$. It leads to an exponential factor $e^{-iz_n t}$ multiplied by the 
pre-exponential function $F_t^n={1\over 2\pi}\int_0^\infty e^{-xt}\delta f(z_n-ix)$, where 
$\delta f(E)=f(E+0^+)-f(E-0^-)$ denotes the jump of the integrand across the branch cut.
We start with the contribution from a {\it branching pole} $z_n\equiv z_k^p$.
Here, one can approximately replace $x\rightarrow 1/t$ in all logarithmic functions and
take the average value $\bar{P}_k(z_k^p-i/t)$ and $\bar{Z}'(z_k^p-i/t)$ across the branch cut.
The jump over the branch cut is dominated by a delta function from the resolvent 
$\delta{1\over E-\lambda_k(E)}\approx 2\pi\delta(x)$. This gives the contribution 
$\rho^{k,\text{p}}_t$ from the branching pole $z_k^p$
\begin{equation}
\label{eq:bp}
\rho^{k,\text{p}}_t \approx e^{-iz_k^p t}\,\bar{P}_k(z_k^p-{i\over t})\,
\bar{Z}'(z_k^p-{i\over t})\,\rho_{t=0}\,.
\end{equation} 
We obtain a logarithmic scaling of the pre-exponential function $F^{k,p}_t$, 
which follows from the scaling behavior of $\bar{P}_k(E)$ and $\bar{Z}'(E)$ around the 
fixed point $z_k^p$. For {\it intermediate times}, where one can expand the logarithmic scaling 
perturbatively in $\alpha\ln(|z_n|t)\ll 1$, one obtains the weak-coupling
expansion of Ref.~\onlinecite{Pletyukhov_Schuricht_Schoeller_10}. However, for {\it long times}, the 
correct scaling behavior has to be determined from a systematic expansion of the full 
solution of the RG equations around the fixed points. For $k=\text{st}$ and $t\rightarrow\infty$
one obtains the stationary distribution $\rho_{\text{st}}=|\bar{x}_{\text{st}}(0)\rangle$ from
Eq.~\eqref{eq:bp}, with $\tilde{L}_\Delta(0)|x_{\text{st}}(0)\rangle=0$.\cite{SM,comment_0}

For a {\it branching point} $z_n$, it is more convenient to start from Eq.~\eqref{eq:pi} and write 
for the jump across the branch cut $\delta\Pi=\Pi_+(\delta L)\Pi_-$ with $\Pi_\pm=\Pi(z_n-ix\pm 0^+)$. 
Using $\Pi_\pm=\tilde{\Pi}_\pm Z_\pm^\prime$, neglecting terms of $O(\delta L)^3$, and approximating
$x\rightarrow 1/t$ in the logarithmic functions, we obtain after some straightforward manipulations the
contribution $\rho_t^{n,\text{b}}$ from the branching point $z_n$:
\begin{eqnarray}
\nonumber
\rho_t^{n,\text{b}} &\approx& e^{-iz_n t}\sum_{k,k'\ne n}{1\over 2\pi}\int_0^\infty dx\,e^{-xt}\\
\label{eq:bc}
&& \hspace{-1.5cm}
\times\bar{P}_k^n \bar{Z}^{\prime n}
{\delta L(z_n-ix) \over (z_n-ix-\bar{\lambda}^n_k)(z_n-ix-\bar{\lambda}^n_{k'})}
\bar{P}_{k'}^n \bar{Z}^{\prime n}\rho_{t=0}\,,
\end{eqnarray}
where $\bar{A}^n=\bar{A}(z_n-i/t)$. For times $t\gg 1/|z_n-z_{k,k'}^p|$, 
the argument $x\sim 1/t$ in the denominators can usually be neglected.\cite{comment_1}
In this case and for {\it intermediate times}, where logarithmic scaling is unimportant, we 
obtain the power law $F^{\text{n,b}}_t\sim 1/t^{1+r}$, where the exponent $r$ follows from the scaling
of the jump of the Liouvillian across the branch cut $\delta L(z_n-ix)\sim x^r$. We therefore set
up an equation for ${d\delta L\over dE}(z_n-ix)$ and solve this RG equation for 
$x\rightarrow 0$. For example, we find $r=0$ for quantum dots in the charge fluctuation regime and
$r=1$ for the Kondo model and the Ohmic spin boson model.\cite{SM} For {\it long times}, one has to
consider in addition the logarithmic corrections from the RG flow 
close to the fixed points.

{\it Ohmic spin boson model.}--- We now apply our flexible method to the Ohmic spin boson model 
at zero bias where a bosonic reservoir $H_{\text{res}}=\sum_q \omega_q a_q^\dagger a_g$ is coupled 
to a two-level system with tunneling $\Delta$, described by the Hamiltonian 
$H=-{\Delta\over 2}\sigma_x$. The coupling is given by 
$V={1\over 2}\sigma_z\sum_q g_q(a_q+a_q^\dagger)$ with ohmic spectral density 
$J(\omega)=\pi\sum_q g_q^2\delta(\omega-\omega_q)=2\pi\alpha\omega\theta(\omega)\rho(\omega)$,
where $\rho(\omega)=D^2/(D^2+\omega^2)$ is some high-energy cutoff function.
We first summarize our results and compare them to previous works.
For weak damping $\alpha\ll 1$, we find three nonzero poles at $z_0=-i\Gamma$ and
$z_\pm=\pm\Omega-i\Gamma/2$, with the effective tunneling $\Omega=\Delta({\Omega\over D})^\alpha$
and $\Gamma=\pi\alpha\Omega$. In leading order truncation it turns out that no
branching poles appear. Therefore, according to the general expressions (\ref{eq:bp}) and 
(\ref{eq:bc}), our results in the intermediate and long time regime $t\gg 1/\Omega$ 
can be written as 
\begin{equation}
\label{eq:rho_tot}
\rho_t = \rho_{\text{st}} + \sum_{n=0,\pm}(F_t^{n,p} + F_t^{n,b})e^{-iz_n t}\,,
\end{equation}
with
\begin{eqnarray}
\label{eq:st}
(\rho_{\text{st}})_{\sigma\sigma} &=& {1\over 2}\quad,\quad
(\rho_{\text{st}})_{\sigma,-\sigma} = {\Omega\over 2\Delta}\\
\label{eq:1p}
F_t^{0,p} &=& {\Omega\over\Delta}
\left(\begin{array}{cc}0 & 0 \\ -1 & \Omega/\Delta  \end{array}\right)
\otimes\tau_+ \,\,\rho_{t=0}\,,\\
\label{eq:sig_p}
F_t^{\sigma,p} &=& {1\over 2}
\left(\begin{array}{cc}1 & \sigma\Omega/\Delta \\ \sigma\Omega/\Delta  & (\Omega/\Delta)^2  \end{array}\right)
\otimes\tau_- \,\,\rho_{t=0}\,,\\
\label{eq:1b}
F_t^{0,b} &=& -{2\alpha\over(\Omega t)^2}
\left(\begin{array}{cc}1 & 0 \\ 0 & 0  \end{array}\right)
\otimes\tau_- \,\,\rho_{t=0}\,,\\
\label{eq:sig_b}
F_t^{\sigma,b} &=& -{\alpha f_t\over(\Delta t)^2}
\left(\begin{array}{cc}0 & 0 \\ 0 & 1  \end{array}\right)
\otimes\tau_+ \,\,\rho_{t=0}\,,
\end{eqnarray}
where $\sigma=\pm$, $\tau_\pm={1\over 2}(1\pm\sigma_x)$, and
\begin{equation}
\label{eq:f_t}
f_t=\left\{[1+\alpha\ln(\Omega t)][1-\ln(1+\alpha\ln(\Omega t))]\right\}^{-2}\,.
\end{equation}
In Liouville space we ordered the four possible
states as $++,--,+-,-+$, where $\pm$ denote the two local spin states. For two $2\times 2$-matrices
$A$ and $B$, we defined the $4\times 4$-matrix 
\begin{equation*}
A\otimes B\equiv\left(\begin{array}{cc}A_{11}B & A_{12}B \\ A_{21}B & A_{22}B  \end{array}\right).
\end{equation*}
The results \eqref{eq:rho_tot}--\eqref{eq:sig_b} can also be obtained from Born\cite{DiVincenzo_05}
and the self-consistent Born\cite{Slutskin_etal_11} approximation, but the unrenormalized 
tunneling $\Omega\rightarrow\Delta$ appears and the pre-exponential functions can only be
calculated up to $O(\alpha)$, i.e., the logarithmic function $f_t$ in Eq.~\eqref{eq:sig_b} is missing.\cite{comment_2} This correction can only be obtained from a resummation 
of all leading logarithmic divergences at low energies. This poses the question of why a 
similar logarithmic correction is not obtained in Eq.~\eqref{eq:1b}. If one compares our
result to the NIBA approximation,\cite{Leggett_87,Weiss_12} which
discusses only the time dynamics of $\langle\sigma_z\rangle_t$
(for $\langle\sigma_{x,y}\rangle_{t=0}=0$ and $\langle\sigma_z\rangle_{t=0}=1$) and reads
\begin{equation}
\label{eq:niba}
\langle\sigma_z\rangle_t\,=\,e^{-{\Gamma\over 2}t}\cos(\Omega t)\,-\,
{2\alpha\over(\Omega t)^{2-2\alpha}}\,,
\end{equation}
one finds, besides the missing exponential part $e^{-\Gamma t}$ in the second term,\cite{comment_3}
a different power law exponent than the one predicted by our result \eqref{eq:1b}.
Below we will show that there is a subtle reason why all leading logarithmic divergencies 
cancel out in Eq.~\eqref{eq:1b}, which is due to the fact that the scaling of the 
vertex and the $Z$-factors is completely different around the fixed points compared with the scaling at
high energies. Our results are further substantiated by bare perturbation theory up to $O(\alpha^2)$
\cite{comment_4}, which confirms that there are no logarithmic terms $\sim\alpha^2\ln(\Omega t)$ in the time   
dynamics, consistent with Eqs.~\eqref{eq:1b}--\eqref{eq:f_t} [note that $f_t=1+O(\alpha^2\ln^2(\Omega t))$].

We now sketch the derivation of our results. First we note that the full effective Liouvillian is
decomposed as $L=L_\Delta(E)+E L'(E) +L^s$, where
\begin{equation*}
L^s=i\pi\alpha\Delta\left(\begin{array}{cc}0 & 0 \\ 1 & 0  \end{array}\right)\otimes\tau_+
\end{equation*}
is the part which arises from integrating out the symmetric part of the reservoir contractions.
The other parts can be parametrized in the following way:
\begin{align*}
\tilde{L}_\Delta &= \sum_\sigma
\left(\begin{array}{cc}0 & 0 \\ 0 & -i\Gamma_\sigma  \end{array}\right)\otimes\tau_\sigma
+\left(\begin{array}{cc}0 & \Delta \\ Z_-\Delta & 0  \end{array}\right)\otimes\tau_-,
\\
Z' &= \sum_\sigma\left(\begin{array}{cc}1 & 0 \\ 0 & Z_\sigma  \end{array}\right)\otimes\tau_\sigma.
\end{align*}
As a consequence, the full propagator can be written as $\Pi=\tilde{\Pi}Z'(1+L^s/E)$, with 
$\tilde{\Pi}=1/(E-\tilde{L}_\Delta)$. In leading order, $\Gamma_-\lesssim \alpha^2\sqrt{Z_-}\Delta$ 
can be neglected and the 
correction from $L^s$ influences only the stationary state (\ref{eq:st}) and the 
pole contribution (\ref{eq:1p}). The Liouvillian $\tilde{L}_\Delta$ has four eigenvalues 
$\lambda_{\text{st}}=0$, $\lambda_0=-i\Gamma_+$ and, neglecting $\Gamma_-$,
$\lambda_\pm=\pm\sqrt{Z_-}\Delta$, with corresponding projectors given by 
\begin{align*}
P_{\text{st}} Z' &= \left(\begin{array}{cc}1 & 0 \\ 0 & 0  \end{array}\right)\otimes\tau_+,
\\
P_0 Z' &= \left(\begin{array}{cc}0 & 0 \\ 0 & Z_+  \end{array}\right)\otimes\tau_+,
\\
P_\pm Z' &= {1\over 2}\left(\begin{array}{cc}1 & \pm\sqrt{Z_-} \\ \pm\sqrt{Z_-} & Z_- 
\end{array}\right)\otimes\tau_-.
\end{align*}
The jump of the Liouvillian across the branch cuts
needed for the evaluation of (\ref{eq:bc}) is parametrized as 
\begin{equation*}
\delta L= \sum_\sigma\left(\begin{array}{cc}0 & 0 \\ 0 & -i\delta\gamma_\sigma 
\end{array}\right)\otimes\tau_\sigma.
\end{equation*}
In leading order, we obtain the following RG equations for $Z_\sigma$, $\Gamma_+$ and $\delta\gamma_\sigma$:\cite{SM}
\begin{eqnarray}
\label{eq:rg_Z}
{dZ_+\over dE}&=&\alpha\tilde{g}\sum_\sigma { Z_+\over E-\lambda_\sigma}\,\,\,,\,\,\,
{dZ_-\over dE}=\alpha\tilde{g}{2Z_-\over E-\lambda_0}\\
\label{eq:rg_gam+}
{d\Gamma_+\over dE}&=&i\alpha\tilde{g}\sum_\sigma {\lambda_0-\lambda_\sigma \over E-\lambda_\sigma}
\\
\label{eq:rg_delta_gam_+}
&&\hspace{-1.2cm}
{d\delta\gamma_+\over dE}=
2\pi\theta(x) \alpha g^2 Z_-\quad\text{for }\,E=z_\sigma-ix\\
\label{eq:rg_delta_gam_-}
&&\hspace{-1.2cm}
{d\delta\gamma_-\over dE}=
4\pi\theta(x) \alpha g^2 Z_+\quad\text{for }\,E=z_0-ix\\
\label{eq:rg_g}
{dg\over dE}&=&\alpha g\tilde{g}\sum_\sigma {1\over \lambda_\sigma}\ln(-i(E-\lambda_\sigma))
\end{eqnarray}
The RG equations are coupled to the renormalization of the vertex function $\tilde{g}=Z_+Z_- g^2$.
The initial conditions at $E=iD$ are $Z_\sigma=1$, $\Gamma_+=\delta\gamma_\pm=0$, and $g^2=1$.
The solution of these RG equations is very different for high and low energies.
For energies $|E|\gg \Omega$ we find $Z_\pm\approx({-iE\over D})^{2\alpha}$, $\tilde{g}\approx 1$,
and $\Gamma_+\approx 0$. According to our general analysis, this gives rise to the
universal shorttime behavior 
\begin{equation*}
\rho_t=\left(\begin{array}{cc}1 & 0 \\ 0 & ({1\over Dt})^{2\alpha}\end{array}\right)
\otimes\mathbbm{1}\,\rho_{t=0},
\end{equation*}
which agrees with previous predictions.\cite{Leggett_87,Weiss_12} In contrast, around the
fixed points we find a different scaling. Close to $z_0=-i\Gamma$, we get 
$Z_+\approx ({\Omega\over\Delta})^2$, 
$Z_-\approx ({\Omega\over\Delta})^2 (1-2\alpha\ln{-i(E-z_0)\over \Omega})^{-1}$,
$\Gamma_+\approx\Gamma$, $g^2 \approx ({\Delta\over\Omega})^4$, and 
$\delta\gamma_-(z_0-ix)\approx-i4\pi\alpha({\Delta\over\Omega})^2 x\theta(x)$. 
Close to $z_\sigma=\sigma\Omega-i\Gamma/2$, we obtain 
$Z_+\approx ({\Omega\over\Delta})^2 (1-\alpha\ln{-i(E-z_\sigma)\over \Omega})^{-1}$,
$Z_-\approx ({\Omega\over\Delta})^2$, 
$\Gamma_+\approx\Gamma/2+i\sigma\Omega\ln(1-\alpha\ln{-i(E-z_\sigma)\over\Omega})$, 
$g^2 \approx ({\Delta\over\Omega})^4$, and 
$\delta\gamma_+(z_\sigma-ix)\approx -i2\pi\alpha({\Delta\over\Omega})^2 x\theta(x)$. Inserting 
these results in the general expressions (\ref{eq:bp}) and (\ref{eq:bc}), one obtains the results 
(\ref{eq:rho_tot}-\ref{eq:f_t}) using straightforward algebra. As we can see, the most 
important point is that the vertex function $\tilde{g}=Z_+Z_-g^2$ is approximately a constant for 
high energies, whereas, at low energies, $g$ is nearly a constant. If one makes a mistake and
takes $\tilde{g}=1$ for {\it all} energies, one would obtain the scaling 
$\delta\gamma_-(z_0-ix)\sim x^{1-2\alpha}\theta(x)$, which produces the incorrect NIBA result
$F_t^{0,b}\sim 1/(\Omega t)^{2-2\alpha}$. On the other hand, if one neglects the 
renormalization of $g$ and takes it as a constant, one obtains the correct scaling for long times but
the scaling of the $Z$-factors at high energies will change with the consequence of an incorrect
description of the dynamics at short times. Therefore, for a proper description of the 
time dynamics on {\it all} time scales, it is crucial to take the vertex renormalization 
into account.

{\it Summary and outlook.} We have shown that the long-time behavior of open quantum
systems involves logarithmic corrections which are generically quite different from those 
at short and intermediate times. For weak-coupling problems we propose a perturbative 
RG method in Liouville space with a complex flow parameter where these corrections can be 
calculated from a systematic expansion around the fixed points. We applied the theory to
the Ohmic spin boson model and found indeed a different long-time behavior than previously
predicted. Moreover, since certain terms of the RG flow are only cut off by the scale $1/t$ of the 
inverse time but not by decay rates $\Gamma$: it is not guaranteed for all models that the RG flow stays
in the weak-coupling regime close to the fixed points irrespective the size of $\Gamma$.
A prominent example of such behavior is the antiferromagnetic nonequilibrium Kondo model at bias
voltage $V$ much larger than the Kondo temperature $T_K$, which
has been shown to be a weak-coupling problem for the determination of stationary quantities.
\cite{Rosch_etal_01_03,Glazman_Pustilnik_05,Kehrein_05,Schoeller_Reininghaus_09} However, by
using the E-RTRG method of Ref.~\onlinecite{Pletyukhov_Schoeller_12}, it turns out that the 
renormalized vertices become strong for $E$ close to the fixed points $z_n=nV-i\Gamma$, 
although $\Gamma$ is much larger than the Kondo temperature $T_K$, whereas they stay small for 
$E$ close to zero (which sets the point to determine stationary quantities). As a consequence,
the long-time behavior can not be calculated from a weak-coupling analysis but
a strong-coupling analysis is needed, which goes 
beyond the perturbative RG method presented in this work and poses new interesting problems
for the future. Other interesting situations arise when one of the poles $z_k^p$ is equal to zero
and is a branching pole, i.e., when there is no exponential decay but a nontrivial preexponential
function with logarithmic scaling. This happens typically for problems with quantum critical 
behavior or for reservoirs with a nonanalytic spectral density in the limit $D\rightarrow\infty$, e.g.,
for multichannel Kondo or sub-Ohmic spin boson models. Such systems are of particular interest since the
long-time behavior is no longer suppressed by the exponential decay and the logarithmic 
scaling behavior becomes more visible. 
 
This work was supported by the DFG via FOR 723. We thank V. Meden for valuable discussions.

\clearpage
\onecolumngrid

\centerline{\large\bf Supplementary Material}
\vspace{.3cm}

We here present technical details of the general derivation of the RG equations
and the application to the ohmic spin boson model. We start with a short summary
of the E-flow scheme of real-time renormalization group (called E-RTRG henceforth).

\section{The E-RTRG method} 

In the supplementary material of Ref.~[11] 
it has been shown how to derive RG equations for the effective Liouvillian $L(E)$ 
by using the Laplace variable $E$ as flow parameter. For 
quantum dots in the spin/orbital fluctuation regime, which are coupled to fermionic reservoirs with
a flat d.o.s., like e.g. the Kondo model, the following RG equations have been derived in lowest order
\begin{align}
\label{eq:L_rg_kondo}
{\partial^2\over \partial E^2}L(E)\quad &= \quad
{1\over 2}
\begin{picture}(10,10)
\put(5,-7){\includegraphics[height=0.7cm]{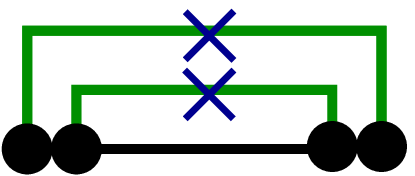}}
\end{picture}
\hspace{2cm}
=\,{1\over 2}\int d\bar{\omega}_1\int d\bar{\omega}_2 \,f'(\bar{\omega}_1)f'(\bar{\omega}_2)
G_{12}(E)\Pi(E_{12}+\bar{\omega}_{12})G_{\bar{2}\bar{1}}(E_{12})
\\ \nonumber \\ 
\label{eq:G_rg_kondo}
{\partial\over \partial E}G_{12}(E)\quad &= \quad
-
\begin{picture}(10,10)
\put(5,-12){\includegraphics[height=1cm]{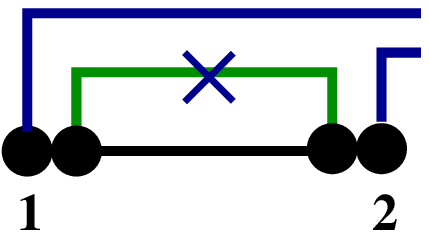}}
\end{picture}
\hspace{2cm}-\,(1\leftrightarrow 2)
\quad=\quad
-\int d\bar{\omega}_3 \,f'(\bar{\omega}_3)
G_{13}(E)\Pi(E_{13}+\bar{\omega}_{3})G_{\bar{3}\bar{2}}(E_{13})\,-\,(1\leftrightarrow 2)
\end{align}
Here, the filled double-circle represents the effective vertex $G_{12}(E)$ at zero frequencies
$\omega_1=\omega_2=0$ which is defined by all connected diagrams with two free reservoir lines. 
The index $1=\eta\alpha\sigma$ with $\eta=\pm$ characterizes the 
reservoir field operators $a_{+\alpha\sigma}= a^\dagger_{\alpha\sigma},
a_{-\alpha\sigma}= a_{\alpha\sigma}$, where $\alpha$ is the reservoir and
$\sigma$ the spin index. The vertex has already been averaged over the Keldysh indices and
its logarithmic frequency dependence has been neglected in leading order.
The black horizontal lines between the vertices represent the full effective
propagator of the local system $\Pi_{1\dots n}=\Pi(E_{1\dots n}+\bar{\omega}_{1\dots n})$
with $\Pi(E)={1\over E-L(E)}$. The indices of the energy argument refer to
the green reservoir contractions running over this propagator, where, for each
contraction, the index has to be taken from the left vertex which is connected to this contraction.
This are precisely the same indices which have to be used to determine the energy
argument for the vertex right to this propagator. 
We have defined $E_{1\dots n}=E+\bar{\mu}_{1\dots n}=E+\bar{\mu}_1+\dots \bar{\mu}_n$,
and $\bar{\omega}_{1\dots n}=\bar{\omega}_1+\dots \bar{\omega}_n$, with
$\bar{\mu}_i=\eta_i\mu_{\alpha_i}$ and $\bar{\omega}_i=\eta_i\omega_i$. Here, $\mu_\alpha$
is the chemical potential of reservoir $\alpha$ and $\omega$ is the single-particle energy of the
reservoir state relative to the chemical potential. The green lines connecting the
vertices represent the reservoir contractions. Including the Keldysh indices they are given by
$\gamma_{11'}^{pp'}(\bar{\omega},\bar{\omega'})=\delta_{1\bar{1}'}\delta(\bar{\omega}+\bar{\omega}')
\gamma^{p'}(\bar{\omega})$, where 
$\gamma^{p'}(\bar{\omega})=p'f(p'\bar{\omega})=f(\bar{\omega})-{1\over 2}+{p'\over 2}$, 
$1=\eta\alpha\sigma$, $\bar{1}'=-\eta'\alpha'\sigma'$, and $f(\omega)={1\over e^{\omega/T}+1}$ 
denotes the Fermi function at temperature $T$. Since the cross at each line denotes the
frequency derivative only ${d\over d\bar{\omega}}\gamma^{p'}(\bar{\omega})=f'(\bar{\omega})$ 
is needed and there is no explicit dependence on the Keldysh indices. This is the reason
why only the vertices averaged over the Keldysh indices appear in the RG equations. 
However, for the perturbative determination of the initial condition it is needed to take the 
symmetric part ${p'\over 2}$ of the reservoir contraction into account.
Symmetry factors ${1\over n!}$ arising from the diagrammatic rules (when two vertices are 
connected by $n$ equivalent lines) are explicitly quoted in (\ref{eq:L_rg_kondo}).
Finally, to calculate the frequency integrals at finite temperature, the following approximate 
form of the propagator has been proposed
\begin{equation}
\label{eq:prop}
\Pi(E+\bar{\omega})\,\approx\,{1\over \bar{\omega} + \chi(E)}\,Z(E) \quad,
\end{equation}
where $Z(E)={1\over 1-{\partial\over \partial E}L(E)}$ is the $Z$-factor superoperator and
$\chi(E)=Z(E)(E-L(E))$. At zero temperature, this approximation is not needed since
$f'(\bar{\omega})=-\delta(\bar{\omega})$ for $T=0$.

Obviously, the frequency integrals are well-defined in the wide band limit $D\rightarrow\infty$, 
i.e. we have obtained universal RG equations. This is the reason why two energy derivatives are
needed for the RG of the Liouvillian, which is typical for problems with spin/orbital fluctuations.
If the same formalism is applied to quantum dots in the charge fluctuation regime, where the
dot states are coupled via tunneling vertices to reservoirs with a flat d.o.s., a single 
derivative is sufficient for convergence and one obtains in leading order the RG equation
\begin{align}
\label{eq:L_rg_dot}
{\partial\over \partial E}L(E)\quad &= \quad
-\,
\begin{picture}(10,10)
\put(5,-7){\includegraphics[height=0.7cm]{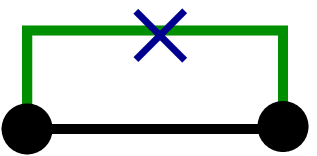}}
\end{picture}
\hspace{1.5cm}
=\,-\int d\bar{\omega}_1 \,f'(\bar{\omega}_1)
G_{1}(E)\Pi(E_{1}+\bar{\omega}_{1})G_{\bar{1}}(E_{1})
\end{align}
Here, in leading order, the unrenormalized vertices can be taken.

For the ohmic spin boson model, where a local $2$-level system is coupled linearly to
an ohmic bosonic bath, the situation is different since the reservoir contractions 
are dressed by the spectral density of the couplings, which is linear in frequency. This
means that the frequency integral in (\ref{eq:L_rg_dot}) becomes logarithmically
divergent and a second derivative w.r.t. the Laplace variable is needed. To show how
the reservoir contractions have to be determined we consider a bosonic bath
$H_{res}=\sum_q\omega_q a^\dagger_q a_q$ with $\omega_q>0$ and a linear
coupling of the form $V=\hat{\gamma}\sum_q g_q(a_q+a_q^\dagger)$, where $\hat{\gamma}$ is 
a generic local operator (i.e. acts only on the states of the local quantum system) and $g_q$ 
denotes the coupling between the local system and mode $q$ of the bath. For an ohmic bath, 
the couplings are characterized by the spectral density 
\begin{equation}
\label{eq:spectral_density}
J(\omega)\,=\,\pi\sum_q g_q^2 \delta(\omega-\omega_q)\,=\,2\pi\,\alpha\,\omega\,\theta(\omega)\,\rho(\omega)
\end{equation}
where $\rho(\omega)={D^2\over D^2+\omega^2}$ is some high-energy cutoff function, which here is
chosen as a Lorentzian for convenience. $\alpha$ is a dimensionless coupling constant characterizing the
damping, which is assumed to be small $\alpha\ll 1$. Using the notation
$a_\eta(\omega)=\sum_q a_{\eta q}\delta(\omega-\omega_q)g_q$, each reservoir contraction
can be expressed via the following average w.r.t. to the canonical distribution of the bosonic bath
\begin{align}
\label{eq:res_contr}
\langle a_\eta(\omega)a_{\eta'}(\omega') \rangle \,=\,
\delta_{\eta,-\eta'}\,\delta(\omega-\omega')\,\eta \,n(\eta\omega)\,{1\over\pi}J(\omega)
\end{align}
where $\langle a_{\eta q}a_{\eta'q'} \rangle = \delta_{\eta,-\eta'}\delta_{qq'}\eta n(\eta\omega_q)$
has been used, with the Bose function $n(\omega)={1\over e^{\omega/T}-1}$. Following 
Ref.~[8], 
this leads to the following contraction between the reservoir 
field operators in Liouville space
\begin{align}
\nonumber
\gamma^{pp'}_{11'}(\bar{\omega},\bar{\omega}')\,&=\,\delta_{1\bar{1}'}\,\delta(\bar{\omega}+\bar{\omega}')
\,{1\over\pi}J(\eta\bar{\omega})\,\eta\, p'\,n(p'\bar{\omega})\\
\label{eq:res_contr_liouville}
&=\,\delta_{1\bar{1}'}\,\delta(\bar{\omega}+\bar{\omega}')\,
2\alpha \,p'\,\bar{\omega}\,n(p'\bar{\omega})\,\theta(\eta\bar{\omega})\,\rho(\bar{\omega})
\end{align}
where the indices $1\equiv\eta$ and $\bar{1}\equiv -\eta$ involve only the creation/annihilation 
index and we have used
$\rho(\omega)=\rho(-\omega)$. Since the local vertex operator $\hat{\gamma}$ is independent of
$\eta$, the contraction (\ref{eq:res_contr_liouville}) can be averaged over $\eta$ and $\eta'$,
which gives 
\begin{align}
\label{eq:res_contr_average}
\gamma^{pp'}(\omega,\omega')\,\equiv\,\sum_{11'}\gamma^{pp'}_{11'}(\omega,\omega')\,=\,
\delta(\omega+\omega')\,\gamma^{p'}(\omega)
\quad,\quad
\gamma^{p'}(\omega)\,=\,2\alpha \,p'\,\omega\,
n(p'\omega)\,\rho(\omega)
\end{align}
In the wide band limit $D\rightarrow\infty$ we omit $\rho(\omega)$. Using $n(-\omega)=-(1+n(\omega))$,
we can split the Bose function via $n(\omega)= -{1\over 2}+(n(\omega)+{1\over 2})$ in symmetric and
antisymmetric part, and obtain for (\ref{eq:res_contr_average})
\begin{align}
\label{eq:res_contr_final}
\gamma^{p'}(\omega)\,=\,\gamma_s^{p'}(\omega)\,+\,\gamma_a(\omega)\quad,\quad
\gamma_s^{p'}(\omega)\,=\,-\alpha \,p'\,\omega\quad,\quad
\gamma_a(\omega)\,=\,\alpha\,\omega\,(2n(\omega)+1)
\end{align}
This gives for the derivatives of the symmetric part
\begin{align}
\label{eq:res_contr_s}
{d\over d\omega}\gamma_s^{p'}(\omega)\,=\,
-\alpha \,p'\quad,\quad
{d^2\over d\omega^2}\gamma_s^{p'}(\omega)\,=\,0\quad.\quad
\end{align}
Most importantly, the first derivative is frequency independent and gives no contribution 
to the RG diagrams (see below). This is the reason why only the 
vertices averaged over the Keldysh indices appear in the RG. For the special
case of zero temperature, we get $n(\omega)=-\theta(-\omega)$, and the antisymmetric part of the 
contraction together with its derivatives reads
\begin{align}
\label{eq:res_contr_T=0}
\gamma_a(\omega)\,=\,\alpha\,|\omega|\quad,\quad
{d\over d\omega}\gamma_a(\omega)\,=\,\alpha\,\text{sign}(\omega)\quad,\quad
{d^2\over d\omega^2}\gamma_a(\omega)\,=\,2\,\alpha\,\delta(\omega)\quad.\quad
\end{align}
Using this form of the contractions, the universal RG equations 
follow from the formalism of Ref.~[11] 
in leading order as
\begin{align}
\label{eq:L_rg_spinboson}
{\partial^2\over \partial E^2}L(E)\quad &= \quad
\begin{picture}(10,10)
\put(5,-7){\includegraphics[height=0.7cm]{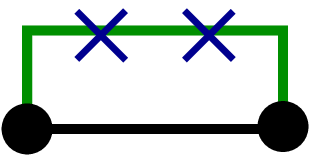}}
\end{picture}
\hspace{2cm}
=\,\int d\omega \,{d^2\gamma_a\over d\omega^2}
G(E)\Pi(E+\omega)G(E)
\\ \nonumber \\ 
\label{eq:G_rg_spinboson}
{\partial\over \partial E}G(E)\quad &= \quad
-
\begin{picture}(10,10)
\put(5,-7){\includegraphics[height=0.8cm]{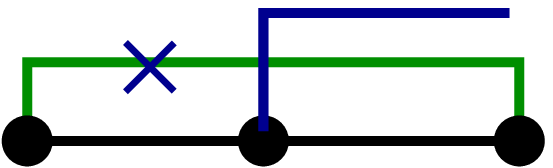}}
\end{picture}
\hspace{3cm} = \quad
-\int d\omega \,{d\gamma_a\over d\omega}
G(E)\Pi(E+\omega)G(E)\Pi(E+\omega)G(E)
\end{align}
Here, the vertex has no further reservoir indices (only for the case of several reservoirs
and different coupling operators $\hat{\gamma}$ to the reservoirs, the reservoir index has
to be retained). Furthermore, the energy argument is just the Laplace variable $E$ since there is
no chemical potential in the bosonic reservoir (we 
allow only for energy exchange with the environment). The frequency integrals are convergent
and we note that the derivatives of the symmetric part of the contraction do not contribute 
in both RG diagrams. To see this for the RG diagram (\ref{eq:G_rg_spinboson}) of the vertex
renormalization, one closes the integration contour in the upper half of the complex plane and 
obtains zero since the derivative of the symmetric part is frequency independent and the 
propagator is an analytic function in the upper half. However, the symmetric part will give
rise to a perturbative correction in the initial condition, see below.
\\ \\
{\it \bf{Decomposition of the Liouvillian}.}--- 
Due to logarithmic divergencies in the high-energy cutoff $D$, the Liouvillian $L(E)$ will
generically depend on various logarithmic terms $\sim\alpha\ln({D\over E-z_n})$, which are
cut off at high energies by $D$ and at low energies by the singularities $z_n$ of the 
propagator $\Pi(E)$. In addition, since the Liouvillian has the dimension of an energy, 
linear terms in the Laplace variable $E$ can occur. In the universal regime $E\ll D$,
the following decomposition is useful, which explicitly exhibits the slowly varying 
logarithmic parts
\begin{equation}
\label{eq:L_decomposition_1}
L(E)\,=\,L_\Delta(E)\,+\,E\,L'(E)\quad.
\end{equation}
Here, $L_\Delta(E)$ is proportional to some physical energy scale $\Delta$ (like e.g. temperature,
magnetic field, decay rates, chemical potentials, level spacing, etc.) but {\it not} the
Laplace variable $E$, and $L_\Delta(E)$ and $L'(E)$ are slowly varying logarithmic functions 
w.r.t. $E$. With this form we can write the propagator as
\begin{align}
\label{eq:prop_new}
\Pi_{1\dots n}&\,\equiv\,\Pi(E_{1\dots n}+\bar{\omega}_{1\dots n})\,=\,
{1\over E_{1\dots n}+\bar{\omega}_{1\dots n} - L(E_{1\dots n}+\bar{\omega}_{1\dots n})}\\
&\,=\,
{1\over E_{1\dots n}+\bar{\omega}_{1\dots n} - L_\Delta(E_{1\dots n}+\bar{\omega}_{1\dots n})
-(E_{1\dots n}+\bar{\omega}_{1\dots n})L'(E_{1\dots n}+\bar{\omega}_{1\dots n})}\,=\,
{1\over E+\chi_{1\dots n}^\Delta}Z^\prime_{1\dots n}\quad,
\end{align}
where 
\begin{align}
\chi^\Delta_{1\dots n}&=\bar{\omega}_{1\dots n}+\bar{\mu}_{1\dots n}-\tilde{L}^\Delta_{1\dots n} \quad,&
\tilde{L}^\Delta_{1\dots n} &=Z^\prime_{1\dots n}L^\Delta_{1\dots n} \quad,&
Z^\prime_{1\dots n}&={1\over 1-L^\prime_{1 \dots n}} \quad,&
Z^\prime(E)&={1\over 1-L^\prime(E)}\quad,
\end{align}
together with 
\begin{align}
L^\Delta_{1\dots n}&=L_\Delta(E_{1\dots n}+\bar{\omega}_{1\dots n})\quad,&
L^\prime_{1\dots n}&=L^\prime(E_{1\dots n}+\bar{\omega}_{1\dots n})\quad,&
Z^\prime_{1\dots n}&=Z^\prime(E_{1\dots n}+\bar{\omega}_{1\dots n})\quad.
\end{align}
Note that $Z(E)={1\over 1-\partial_EL(E)}$ and $Z'(E)={1\over 1-L'(E)} $ 
are not identical since
\begin{equation}
\label{eq:Z.vs.Z'}
\partial_E L(E)\,=\,L'(E)\,+\,\left\{\partial_E L_\Delta(E) \,+\, E\,\partial_E L'(E)\right\}
\,=\,L'(E)\,+\,O(G^2)\quad.
\end{equation}
Below we will show that $\partial_E L_\Delta(E) + E\,\partial_E L'(E)\sim O(G^2)$, i.e. $Z(E)$ and
$Z'(E)$ are the same in leading order up to $O(G)$.

We introduce the following graphical notations for the lines connecting the vertices
\begin{align}
\label{eq:prop_diagrams}
\Pi_{1\dots n}&\equiv 
\begin{picture}(10,10)
\put(5,-2){\includegraphics[height=0.3cm]{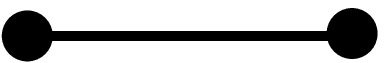}}
\end{picture}
\hspace{2cm}&,\quad\quad
Z'_{1\dots n}&\equiv
\begin{picture}(10,10)
\put(5,-2){\includegraphics[height=0.3cm]{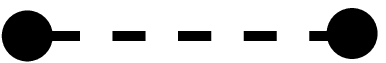}}
\end{picture}
\hspace{2cm}&,\quad\quad
\chi^\Delta_{1\dots n}\Pi_{1\dots n}&\equiv 
\begin{picture}(10,10)
\put(5,-8){\includegraphics[height=0.5cm]{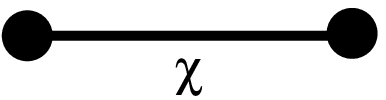}}
\end{picture}
\hspace{3cm},
\end{align}
such that the relation $Z'_{1\dots n}-E\,\Pi_{1\dots n}=\chi^\Delta_{1\dots n}\Pi_{1\dots n}$ can
be written diagrammatically as
\begin{equation}
\label{eq:prop_relation}
\begin{picture}(10,10)
\put(5,-1){\includegraphics[height=0.3cm]{prop_full_2.eps}}
\end{picture}
\hspace{2cm}-\,\, E
\begin{picture}(10,10)
\put(5,-1){\includegraphics[height=0.3cm]{prop_full_1.eps}}
\end{picture}
\hspace{2cm}=\,\,
\begin{picture}(10,10)
\put(5,-7){\includegraphics[height=0.5cm]{prop_full_3.eps}}
\end{picture}
\hspace{3cm}.
\end{equation}
To find RG equations for $L_\Delta(E)$ and $L'(E)$, we consider the three
cases of spin/orbital fluctuations (e.g. Kondo model, Eq.~(\ref{eq:L_rg_kondo})), 
charge fluctuations (e.g. quantum dots, Eq.~(\ref{eq:L_rg_dot})) and energy fluctuations 
(e.g. spin boson model, Eq.~(\ref{eq:L_rg_spinboson})) separately. For charge fluctuations,
the RG equation (\ref{eq:L_rg_dot}) is already of first order, i.e. the Liouvillian
$L(E)\approx L_\Delta(E)$ is in leading order a logarithmic function and we get
\begin{align}
\label{eq:L_rg_firstorder_dot}
\hspace{-2cm}
\text{\underline{Charge fluctuations}:}\hspace{2cm}
{\partial\over \partial E} L_\Delta(E)\quad = \quad
-\,
\begin{picture}(10,10)
\put(5,-7){\includegraphics[height=0.7cm]{L_cross_single.eps}}
\end{picture}
\hspace{2cm},\quad
{\partial\over \partial E} L'(E)\quad = \quad 0
\end{align}
For spin/orbital fluctuations we write the second order differential 
equations (\ref{eq:L_rg_kondo}) formally as
\begin{align}
\label{eq:kondo_zw}
{\partial^2\over \partial E^2}L(E)\quad &= \quad
{1\over 2}
\begin{picture}(10,10)
\put(5,-7){\includegraphics[height=0.7cm]{L_cross_free.eps}}
\end{picture}
\hspace{2cm}
+\quad{\partial\over\partial E}\left\{{1\over 2}
\begin{picture}(10,10)
\put(5,-7){\includegraphics[height=0.7cm]{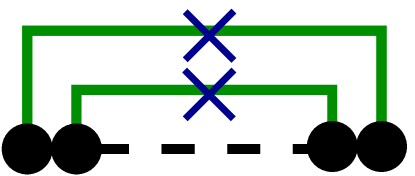}}
\end{picture}
\hspace{1.5cm}\right\}\quad+\quad O(G^3)\quad.
\end{align}
Thereby, the $E$-derivative of the $Z$-factors and vertices in the 
second diagram on the r.h.s. gives rise to terms of $O(G^3)$ and are added for convenience. 
We now identify the r.h.s. of this equation with the expression
\begin{equation}
\label{eq:L_second_derivative}
{\partial^2\over \partial E^2}L(E)\,=\,{\partial\over \partial E}L'(E)
\,+\,{\partial\over \partial E}\left\{{\partial\over \partial E}L_\Delta
\,+\,E{\partial\over \partial E}L'(E)\right\}
\end{equation}
and find 
\begin{align}
\label{eq:L'_rg}
{\partial\over \partial E}L'(E)\quad &= \quad
{1\over 2}
\begin{picture}(10,10)
\put(5,-7){\includegraphics[height=0.7cm]{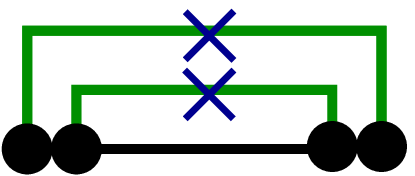}}
\end{picture}
\\ \nonumber \\
\label{eq:partial_L.minus.L'}
{\partial\over \partial E}L_\Delta
\,+\,E{\partial\over \partial E}L'(E)\,&=\,
{\partial\over\partial E}L(E)\,-\,L'(E)\,=\,
{1\over 2}
\begin{picture}(10,10)
\put(5,-7){\includegraphics[height=0.7cm]{SD_L_1_free.eps}}
\end{picture}
\hspace{1.2cm}\quad+\quad O(G^3)\quad.
\end{align}
The second equation shows that the difference between $\partial_E L(E)$ and
$L'(E)$ is indeed of $O(G^2)$ as stated above after Eq.~(\ref{eq:Z.vs.Z'}).
Solving (\ref{eq:partial_L.minus.L'}) for $\partial_E L_\Delta(E)$ and
using (\ref{eq:L'_rg}) and (\ref{eq:prop_relation}) we find in leading order
\begin{align}
\label{eq:L_rg_firstorder_kondo}
\hspace{-2cm}
\text{\underline{Spin/orbital fluctuations}:}\hspace{1cm}
{\partial\over \partial E} L_\Delta(E)\quad = \quad
{1\over 2}
\begin{picture}(10,10)
\put(5,-12){\includegraphics[height=0.9cm]{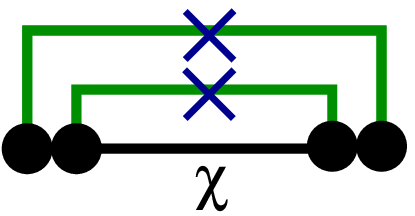}}
\end{picture}
\hspace{2cm},\quad
{\partial\over \partial E} L'(E)\quad = \quad 
{1\over 2}\,
\begin{picture}(10,10)
\put(5,-7){\includegraphics[height=0.7cm]{SD_L_1_full.eps}}
\end{picture}
\end{align}
As required we find that $L_\Delta(E)$ is proportional to some physical
energy scale appearing in $\chi^\Delta_{1\dots n}$. Note that, at finite
temperature, also the frequencies $\bar{\omega}_{1\dots n}$ appear in
$\chi^\Delta_{1\dots n}$, which gives a term proportional to temperature
for $L_\Delta(E)$. 

The same procedure can be applied to the RG equation (\ref{eq:L_rg_spinboson})
for the case of energy fluctuations with the result
\begin{align}
\label{eq:L_rg_firstorder_spinboson}
\hspace{-2cm}
\text{\underline{Energy fluctuations}:}\hspace{1cm}
{\partial\over \partial E} L_\Delta(E)\quad = \quad
\,
\begin{picture}(10,10)
\put(5,-12){\includegraphics[height=0.9cm]{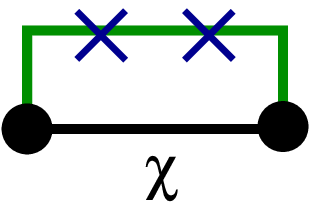}}
\end{picture}
\hspace{2cm},\quad
{\partial\over \partial E} L'(E)\quad = \quad 
\begin{picture}(10,10)
\put(5,-7){\includegraphics[height=0.7cm]{L_cross_spinboson.eps}}
\end{picture}
\end{align}

The RG equations (\ref{eq:L_rg_firstorder_dot}), (\ref{eq:L_rg_firstorder_kondo}) and
(\ref{eq:L_rg_firstorder_spinboson}), together with the vertex renormalization
(\ref{eq:G_rg_kondo}) and (\ref{eq:G_rg_spinboson}) are the final differential equations
to be solved to obtain the Liouvillian $L(E)$ in the form (\ref{eq:L_decomposition_1}).
We note that the choice of the second term in (\ref{eq:kondo_zw}) seems at first
sight not unique since it is of higher order. However, it is not important whether
the construction of the RG equations for $ L_\Delta(E)$ and $L'(E)$ is
unique at a certain truncation order, but the crucial point is that all corrections
to this construction can be shown to be beyond leading order.

With the quantities $L_\Delta(E)$ and $L'(E)$ the propagator $\Pi(E)$ can be expressed as
\begin{equation}
\label{eq:tilde_propagator}
\Pi(E)\,=\,{1\over E-L(E)}\,=\,\tilde{\Pi}(E)\,Z'(E)\quad,\quad
\tilde{\Pi}(E)\,=\,{1\over E-\tilde{L}_\Delta(E)}\,
\end{equation}
with
\begin{equation}
\label{eq:def_tilde_L_Z}
\tilde{L}_\Delta(E)\,=\,Z'(E)\,L_\Delta(E)\quad,\quad 
Z'(E)\,=\,{1\over E-L'(E)}\quad.
\end{equation}
To evaluate the RG equations one needs the frequency dependence of the propagator 
$\Pi(E+\omega)$. This is approximated by neglecting it in leading order within the 
logarithmic parts $\tilde{L}_\Delta$ and $Z'$, i.e. we use
\begin{equation}
\label{eq:prop_frequency}
\Pi(E+\omega)\,\approx\,{1\over E+\omega-\tilde{L}_\Delta(E)}\,Z'(E)\quad.
\end{equation}

Finally, due to the properties $\text{Tr} L(E)=0$ and $L(E)^c=-L(-E^*)$ 
(with the $c$-transform defined by $(L^c)_{s_1 s_2,s_{1'}s_{2'}}=(L_{s_2 s_1,s_{2'}s_{1'}})^*$)
[8], 
we note the following properties for the quantities $\tilde{L}_\Delta(E)$
and $Z'(E)$
\begin{equation}
\label{eq:properties_LZ}
\text{Tr}\,\tilde{L}_\Delta(E)\,=\,0\quad,\quad
\text{Tr}\,Z'(E)\,=\,1\quad,\quad
\tilde{L}_\Delta(E)^c\,=\,-\tilde{L}_\Delta(-E^*)\quad,\quad
Z'(E)^c\,=\,Z'(-E^*)\quad.\quad
\end{equation}
\\ 
{\it \bf{Time evolution}.}--- 
Once the quantities $\tilde{L}_\Delta(E)$ and $Z'(E)$ are known, the time evolution
of the local density matrix $\rho_t$ can be discussed in a straightforward way.
Choosing any initial state $\rho_{t=0}$ at $t=0$, we get in Laplace space the
solution $\rho(E)=\int_0^\infty dt\,e^{iEt}\rho_t=i\Pi(E)\rho_{t=0}$, and in
time space via inverse Laplace transform
\begin{equation}
\label{eq:time_evolution}
\rho_t\,=\,{i\over 2\pi}\,\int_{-\infty+i0^+}^{\infty+i0^+}dE\,e^{-iEt}\,
{1\over E-L(E)}\,\rho_{t=0}\,=\,
{i\over 2\pi}\,\int_{-\infty+i0^+}^{\infty+i0^+}dE\,e^{-iEt}\,
{1\over E-\tilde{L}_\Delta(E)}\,Z'(E)\,\rho_{t=0}\quad,
\end{equation}
The last form of (\ref{eq:time_evolution}) is very helpful for the evaluation of the
energy integral because it explicitly exhibits the slowly varying logarithmic functions
$\tilde{L}_\Delta(E)$ and $Z'(E)$. To identify the singularities of the integrand we
use the spectral decomposition of the Liouvillian $\tilde{L}_\Delta(E)$ in terms of its
eigenvalues $\lambda_k(E)$ and corresponding projectors $P_k(E)$ 
\begin{equation}
\label{eq:spectral}
\tilde{L}_\Delta(E)\,=\,\sum_k\,\lambda_k(E)\,P_k(E)
\end{equation}
Since we deal with a non-hermitian superoperator, we have to distinguish the left and
right eigenvectors, which we denote in Dirac notation by $|x_k(E)\rangle$ and $\langle \bar{x}_k(E)|$
\begin{equation}
\label{eq:eigenvectors}
\tilde{L}_\Delta(E)\,|x_k(E)\rangle\,=\,\lambda_k(E)\,|x_k(E)\rangle\quad,\quad
\langle \bar{x}_k(E)|\,\tilde{L}_\Delta(E)\,=\,\langle \bar{x}_k(E)|\,\lambda_k(E)\quad.\quad
\end{equation}
The eigenvectors fulfill the orthonormalization condition 
$\langle \bar{x}_k(E)|x_{k'}(E)\rangle=\delta_{kk'}$ and the projectors are given by 
$P_k(E)=|x_k(E)\rangle\langle \bar{x}_k(E)|$ with $\sum_k P_k(E)=1$.

Due to the condition $\text{Tr}\tilde{L}_\Delta(E)=0$, we obtain either $\lambda_k(E)=0$
or $\text{Tr}\,|x_k(E)\rangle=0$. Therefore, the Liouvillian has always an eigenvalue zero,
which we characterize by the index $k=\text{st}$ since it corresponds to the stationary state (see below).
The other eigenvalues are numerated by $k=0,\pm 1,\pm 2, \dots$. We get
\begin{eqnarray}
\label{eq:zero_eigenvectors}
\quad\text{Tr}\,|x_{\text{st}}(E)\rangle\,&=&\,\sum_s\,\langle ss|x_{\text{st}}(E)\rangle\,=\,1\quad,\quad
\langle \bar{x}_{\text{st}}(E)|ss\rangle\,=\,1\\
\label{eq:tr_eigenvectors}
\text{Tr}\,|x_k(E)\rangle\,&=&\,\sum_s\,\langle ss|x_k(E)\rangle\,=\,0\quad,\quad
\text{for}\quad k=0,\pm 1,\pm 2,\dots\quad.
\end{eqnarray}
As a consequence we get $P_{\text{st}}(E)=|x_{\text{st}}(E)\rangle\text{Tr}$ and the property
$\text{Tr}L_\Delta(E)=\text{Tr}L'(E)=0$ can also be written as
\begin{equation}
\label{eq:P_st_property}
P_{\text{st}}(E)\,Z'(E)\,=\,P_{\text{st}}(E)\quad,\quad
P_{\text{st}}(E)\,L_\Delta(E)\,=\,0\quad.
\end{equation}
Due to the condition $\tilde{L}(E)^c=-L(-E^*)$, the eigenvalues and projectors occur always 
in pairs (except for $k=0,\text{st}$ where we define $k\equiv -k$) with 
\begin{equation}
\label{eq:pairs}
\lambda_{-k}(E)\,=\,-\lambda_k(-E^*)\quad,\quad
P_{-k}(E)\,=\,P_k(-E^*)^c\quad.
\end{equation}

Using the spectral representation, the time evolution can be written as
\begin{equation}
\label{eq:time_evolution_spectral}
\rho_t\,=\,
{i\over 2\pi}\,\sum_k\,\int_\gamma dE\,e^{-iEt}\,
{1\over E-\lambda_k(E)}\,P_k(E)\,Z'(E)\,\rho_{t=0}\quad,
\end{equation}
where $\gamma$ is an integration contour which encloses the lower half of the
complex plane including the real axis. Poles are located at $E=z_k^p=\lambda_k(z_k^p)$,
where $z_{\text{st}}^p=0$ is a pole at the origin. At {\bf zero temperature}, which we consider
from now on, additional nonanalytic features occur from
branch cuts since $\lambda_k(E)$, $P_k(E)$ and $Z'(E)$ depend
logarithmically via terms $\sim \ln({D\over E-z_n})$ generated by the ultraviolet
divergencies from the high-energy cutoff $D$ (at finite temperature the branch cuts turn
into an infinite number of discrete poles separated by $2\pi T$). From the structure of the
perturbation theory it can be seen that the singularities $z_n$ are associated
with poles of the propagators $\tilde{\Pi}(E_{1\dots n})$, i.e. are located at 
$E_{1\dots n}=z_k^p$, where $E_{1\dots n}=E+\bar{\mu}_{1\dots n}$.
Therefore, the singularities $z_n=z_k^p-\bar{\mu}_{1\dots n}$ are generically given by the 
poles shifted by some linear combination of the chemical potentials of the reservoirs.

Using the general expressions (\ref{eq:time_evolution}) and (\ref{eq:time_evolution_spectral}),
one can discuss the qualitative form of the time evolution in different time regimes. For
{\it short times} $t\ll 1/|z_n|$, we obtain $E\sim 1/t\gg |z_n|$, i.e. the cutoff scales $z_n$ 
in the logarithmic terms are unimportant and can be neglected. Furthermore, in leading order,
we can replace $E\rightarrow 1/t$ in the logarithmic parts. This means that we cut off the
poor man scaling equations for $\tilde{L}_\Delta(E)$ and $Z'(E)$ at the scale $E\sim 1/t$ 
and obtain from (\ref{eq:time_evolution})
\begin{equation}
\label{eq:short_times}
\rho_t\,=\,{i\over 2\pi}\,\int_\gamma dE\,e^{-iEt}\,
{1\over E-\tilde{L}_\Delta(1/t)}\,Z'(1/t)\,\rho_{t=0}
\,=\,e^{-i\tilde{L}_\Delta(1/t)t}\,Z'(1/t)\,\rho_{t=0}\quad.
\end{equation}
Expanding the exponential one finds in leading order that the logarithmic scaling of 
$Z'(1/t)$ at high energies determines the short time behavior.

For {\it intermediate and long times} $t\gtrsim 1/|z_n|$, we have to study the contributions from
the poles and branch cuts in detail. All branch cuts are chosen to point into the 
direction of the negative imaginary axis, i.e. are located at $z_n-ix$ with $x>0$.
This can even be enforced numerically by solving the RG equations along the two paths
$E=z_n+i\Lambda\pm 0^+$ with $\Lambda$ real and initially given by $\Lambda=D$. Since no
singularities are surrounded by the two paths, the RG flow is analytic and can be used
to determine the jump across the branch cut. This is a particular advantage of the
E-RTRG method, which uses a complex flow parameter in Laplace space. Our choice for
the direction of the branch cuts is very convenient since 
$e^{-iEt}=e^{-iz_n t}e^{-xt}$ is exponentially decaying in $xt$, which allows an
analytical evaluation of the branch cut integrals for intermediate and long times.
We start with the contributions from the branch cuts starting at a pole or  
branching pole at $z_k^p$, which we evaluate by using the form
(\ref{eq:time_evolution_spectral}). For the branch cut integral
we set $E=z_k^p-ix\pm 0^+$ and replace in leading order $\lambda_k(E)\rightarrow z_k^p$ 
and the logarithmic function $P_k(E)Z'(E)$ by its average
$\bar{P}_k(z_k^p-ix)\bar{Z}'(z_k^p-ix)$ over the branch cut, where 
$\bar{A}(E)={1\over 2}(A(E+0^+)+A(E-0^+))$. Furthermore, in leading order, we can 
use $x\rightarrow 1/t$ in the logarithmic functions. This gives the contribution
\begin{equation}
\label{eq:branching_pole_zw}
\rho_t^{k,p}\,\approx\,e^{-iz_k^p t}\,{1\over 2\pi}\,\int_{0^-}^\infty dx\,e^{-xt}\,
\left({1\over -ix+0^+}-{1\over -ix-0^+}\right)\,\bar{P}_k(z_k^p-i/t)\,\bar{Z}'(z_k^p-i/t)
\,\rho_{t=0}\quad.
\end{equation}
Using ${1\over -ix+0^+}-{1\over -ix-0^+}=2\pi\delta(x)$, we obtain
\begin{equation}
\label{eq:branching_pole}
\rho_t^{k,p}\,\approx\,e^{-iz_k^p t}\,\bar{P}_k(z_k^p-i/t)\,\bar{Z}'(z_k^p-i/t)\rho_{t=0}
\quad,
\end{equation}
i.e., for $z_k^p=\Omega_k-i\Gamma_k$, an exponential time evolution with oscillation $\Omega_k$
and decay rate $\Gamma_k$, modulated by a logarithmic scaling function. 
For the special term $k=\text{st}$, where $z^p_{\text{st}}=0$, 
$P_{\text{st}}(E)=|x_{\text{st}}(E)\rangle\text{Tr}$ and $P_{\text{st}}(E)Z'(E)=P_{\text{st}}(E)$, 
we get the following contribution to the time evolution
\begin{equation}
\label{eq:branching_pole_zero}
\rho_t^{\text{st},p}\,\approx\,|\overline{x_{\text{st}}}(-i/t)\rangle
\,\xrightarrow{t\rightarrow\infty}\,\rho_{\text{st}}\,=\,|\overline{x_{\text{st}}}(0)\rangle
\quad,
\end{equation}
i.e. we see that for $t\rightarrow\infty$ one always gets the stationary distribution
$\rho_{\text{st}}$ but, if $z_{\text{st}}^p$ is a branching pole, logarithmic corrections 
can occur for the time evolution which do not decay exponentially. 
We note that for the models discussed here, there is no logarithmic term in the 
diagrammatic series involving the pole $z^p_{\text{st}}$. The reason is that 
the projector $P_{\text{st}}$ gives always a regular contribution, provided that the symmetric part 
$\gamma_s^{p'}(\omega)$ of the contraction (\ref{eq:res_contr_final}) is an analytic 
function [8]. 
Providing there is no accidental pole $z_{k\ne\text{st}}^{p}=0$, 
the pole at $E=0$ is isolated and has no attached branch cuts.

The evaluation of a branch cut starting at a branching point $z_n$ which is not a pole is more 
subtle since both $\lambda_k(E)$ and $P_k(E)Z'(E)$ can be discontinuous and cancellations can occur 
between the two contributions. Therefore, it is more convenient to start from the first expression
of (\ref{eq:time_evolution}) involving the propagator $\Pi(E)$. Denoting by 
$\delta A=A_+-A_-$ the jump across the branch and by $\bar{A}={1\over 2}(A_++A_-)$ the
average value, with $A_\pm=A(E\pm 0^+)=\bar{A}\pm{1\over 2}\delta A$, one finds for the 
jump of the propagator expanding in small $\delta L \sim G\bar{L}$
\begin{equation}
\label{eq:jump_propagator}
\delta\Pi(E)\,=\,\Pi_+\,\delta L\,\Pi_-\,=\,{1\over E-\bar{L}}\,\delta L\,{1\over E-\bar{L}}
\,+\,O(\delta L^3)\quad.
\end{equation}
Using $\overline{AB}-\bar{A}\bar{B}={1\over 4}\delta A\delta B$, we get
\begin{equation}
\label{eq:average_propagator}
{1\over E-\bar{L}}\,=\,\overline{{1\over E-L}}+O(\delta L^2)\,=\,
\sum_k\overline{{1\over E-\lambda_k}P_k Z'}+O(\delta L^2)\,=\,
\sum_k{1\over E-\bar{\lambda}_k}\bar{P}_k\bar{Z}'+O(\delta L^2)
\end{equation}
Inserting this in (\ref{eq:jump_propagator}), neglecting $O(\delta L^3)$, and approximating
$E=z_n-ix\rightarrow z_n-i/t$ in the
logarithmic functions $\bar{\lambda}_k$, $\bar{P}_k$ and $\bar{Z}'$, we get  
the following result for the branch cut integral
\begin{equation}
\label{eq:branching_point}
\rho_t^{n,b}\,\approx\,e^{-iz_nt}\,{1\over 2\pi}\,\sum_{kk'\ne n}\,\int_0^\infty dx\,e^{-xt}\,
{1\over z_n-ix-\bar{\lambda}^n_k}\,\bar{P}^n_k\,\bar{Z}^{\prime n}\,\delta L(z_n-ix)\,
{1\over z_n-ix-\bar{\lambda}^n_{k'}}\,\bar{P}^n_{k'}\,\bar{Z}^{\prime n}\,\rho_{t=0}
\quad,
\end{equation}
where $\bar{\lambda}_k^n=\bar{\lambda}_k(z_n-i/t)$, $\bar{P}_k^n=\bar{P}_k(z_n-i/t)$ and
$\bar{Z}^{\prime n}=\bar{Z}'(z_n-i/t)$. We have omitted the cases $k=n$ or $k'=n$ since
we consider a branching point and not a branching pole. Since
$\bar{\lambda}_k^n\sim O(z_k^p)$, we can neglect $x$ in the denominators of the resolvents
for times $t\sim 1/x \gg 1/|z_n-z_{k,k'}^p|$. For special resonant cases, where $z_n$ comes
close to $z_k^p$ or $z_{k'}^p$, one can also define time regimes 
$1/|z_n|\lesssim t\ll 1/|z_n-z_{k,k'}^p|$, where $x$ dominates in the denominators for certain
values of $k$ or $k'$. In any case, to evaluate the integral over $x$, it is necessary to
know the jump of the Liouvillian $\delta L(z_n-ix)$, for which we will derive RG equations
in the following.
\\ \\
{\it \bf{RG equation for $\delta L$}.}--- 
In leading order, the jump $\delta L$ of the Liouvillian at a branch cut with $E=z_n-ix\pm 0^+$
is generated by some propagator $\Pi_{1\dots n}$ in the perturbative expansion, which is resonant, i.e. the
jump of this propagator across the branch cut becomes a $\delta$-function. To tune the branch cut of
each propagator w.r.t. $E$ along the direction of the negative imaginary axis, we first
close all integration contours over the real frequencies $\bar{\omega}$ in the upper half of 
the complex plane, where the only nonanalytical properties are those of the Fermi/Bose-functions 
on the positive imaginary axis. This turns all frequency integrations to
ones along the positive imaginary axis $\int d\bar{\omega}\rightarrow i\int_0^\infty d\bar{\omega}$ 
and the sign-functions of the antisymmetric part of the Fermi/Bose functions have to be replaced 
by their jump on the imaginary axis $\text{sign}(\bar{\omega})\rightarrow 2$. 
A particular resolvent containing the eigenvalue $\lambda_k(E_{1\dots n}+i\bar{\omega}_{1\dots n})$ 
will then become resonant if the condition $z_n=z_k^p-\bar{\mu}_{1\dots n}$ is fulfilled. 
With $E_{1\dots n}=z_n+\bar{\mu}_{1\dots n}-ix\pm 0^+=z_k^p-ix\pm 0^+$, 
we replace approximately $\lambda_k\rightarrow z_k^p$, 
$P_k\rightarrow \bar{P}_k(z_k^p-ix)$ and $Z'\rightarrow \bar{Z}'(z_k^p-ix)$, which gives
for the jump of the propagator the following $\delta$-function
\begin{eqnarray}
\nonumber
\delta\Pi_{1\dots n}\,&=&\,\left({1\over -ix+i\bar{\omega}_{1\dots n}+0^+}-
{1\over -ix+i\bar{\omega}_{1\dots n}-0^+}\right)\,\bar{P}_k(z_k^p-ix)\,\bar{Z}'(z_k^p-ix)\\
\nonumber\\
\label{eq:prop_res}
&=&\,2\pi \delta(\bar{\omega}_{1\dots n}-x)\,\bar{P}_k(z_k^p-ix)\,\bar{Z}'(z_k^p-ix)\quad.
\end{eqnarray}
Since $\bar{\omega}_{1\dots n}>0$, the frequency integrals give only a contribution for $x>0$.
Diagrammatically, we indicate the jump of the propagator by
\begin{align}
\label{eq:prop_jump_diagram}
\delta\Pi_{1\dots n}\quad\equiv\quad 
\begin{picture}(10,10)
\put(5,-6){\includegraphics[height=0.4cm]{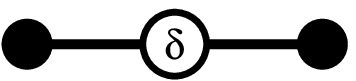}}
\end{picture}
\hspace{3cm}.
\end{align}
The RG equations ${\partial\delta L\over \partial E}(z_n-ix)$ can be obtained by a similar 
technique as the RG equations for ${\partial\over \partial E}L(E)$ and ${\partial^2\over \partial E^2}L(E)$.
We obtain in leading order
\begin{align}
\label{eq:delta_L_rg_dot_diag}
\hspace{-2cm}
\text{\underline{Charge fluctuations}:}&\hspace{2.55cm}
\delta L(E)\quad = \quad
\begin{picture}(10,10)
\put(5,-7){\includegraphics[height=0.7cm]{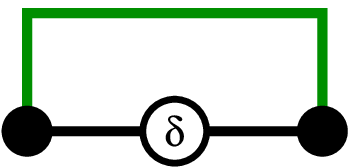}}
\end{picture}
\\
\label{eq:delta_L_rg_kondo_diag}
\hspace{-2cm}
\text{\underline{Spin/orbital fluctuations}:}&\hspace{2cm}
{\partial\over\partial E}\delta L(E)\quad = \quad-\,{1\over 2}
\begin{picture}(10,10)
\put(5,-7){\includegraphics[height=0.7cm]{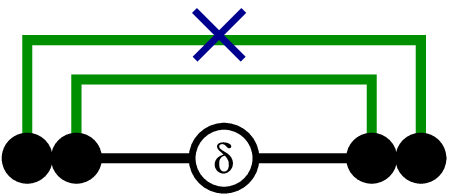}}
\end{picture}
\\
\label{eq:delta_L_rg_spinboson_diag}
\hspace{-2cm}
\text{\underline{Energy fluctuations}:}&\hspace{2cm}
{\partial\over\partial E}\delta L(E)\quad = \quad-
\begin{picture}(10,10)
\put(5,-7){\includegraphics[height=0.7cm]{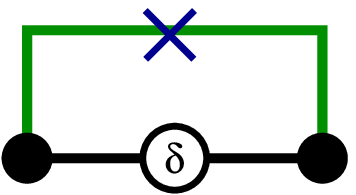}}
\end{picture}
\end{align}
We have chosen the number of derivatives by the criterion that the frequency integrals
on the r.h.s. are convergent.

These equations are nontrivial only on the branches, where $E=z_n-ix$ and $z_n=z_k^p-\bar{\mu}_{1\dots n}$.
Setting this energy argument in the propagators between the vertices, 
we explicitly obtain together with (\ref{eq:prop_res}) (note that we consider zero temperature)
\begin{align}
\nonumber
&\text{\underline{Charge fluctuations}:}\\
\label{eq:delta_L_rg_dot}
&\hspace{4cm}
\delta L(z_n-ix)\,=\,
-2\pi i\,\theta(x)\,\bar{G}_1(z_k^p-\bar{\mu}_1-ix)\,
\bar{P}_k(z_k^p-ix)\,\bar{Z}'(z_k^p-ix)\,\bar{G}_{\bar{1}}(z_k^p-ix)
\\
\nonumber
&\text{\underline{Spin/orbital fluctuations}:}\\
\label{eq:delta_L_rg_kondo}
&\hspace{4cm}
{\partial\over\partial x}\delta L(z_n-ix)\,=\,
-2\pi\,\theta(x)\,\bar{G}_{12}(z_k^p-\bar{\mu}_{12}-ix)\,\bar{P}_k(z_k^p-ix)\,\bar{Z}'(z_k^p-ix)\,\bar{G}_{\bar{2}\bar{1}}(z_k^p-ix)
\\
\nonumber
&\text{\underline{Energy fluctuations}:}\\
\label{eq:delta_L_rg_spinboson}
&\hspace{4cm}
{\partial\over\partial x}\delta L(z_n-ix)\,=\,
-4\pi\alpha \,\theta(x)\,\bar{G}(z_n-ix)\,\bar{P}_k(z_n-ix)\,\bar{Z}'(z_n-ix)\,\bar{G}(z_n-ix)
\end{align}
where $z_n=z_k^p$ in the case of energy fluctuations, and we have replaced all vertices by their average
$\bar{G}$ across the branch cut in case that they are discontinuous. 
The initial condition for the last two equations is $\delta L(z_n)=0$.
Up to the corrections from the logarithmic functions, we obtain $\delta L(z_n-ix)\sim \theta(x)$ for
charge fluctuations and $\delta L(z_n-ix)\sim x\,\theta(x)$ for spin/orbital and energy fluctuations.
Therefore, if $x$ can be neglected in the resolvents of the integrand of (\ref{eq:branching_point}),
we obtain (up to logarithmic corrections) $\rho_t^{n,b}\sim 1/t$ for charge fluctuations and
$\rho_t^{n,b}\sim 1/t^2$ for spin/orbital and energy fluctuations.  

\section{Application to the ohmic spin boson model}
\label{sec:spin_boson}
We now apply the formalism to the ohmic spin boson model at zero bias, defined by
the Hamiltonian $H_{\text{tot}}=H+H_{\text{res}}+V$ with
\begin{equation}
\label{eq:H_spinboson}
H\,=\,-{\Delta\over 2}\sigma_x \quad,\quad
H_{\text{res}}\,=\,\sum_q\omega_q a^\dagger_q a_q \quad,\quad
V\,=\,{1\over 2}\sigma_z\sum_q g_q (a_q+a^\dagger_q) \quad.\quad
\end{equation}
According to (\ref{eq:spectral_density}), we use an ohmic spectral density and the 
reservoir contraction in Liouville space is given at zero temperature by (\ref{eq:res_contr_final}) 
\begin{equation}
\label{eq:contraction_T=0}
\gamma^{p'}(\omega)\,=\,\gamma_s^{p'}(\omega)\,+\,\gamma_a(\omega)\quad,\quad
\gamma_s^{p'}(\omega)\,=\,-\alpha \,p'\,\omega\quad,\quad
\gamma_a(\omega)\,=\,\alpha\,|\omega|\quad.
\end{equation}

To set up the algebra for the Liouvillian $L(E)$ and the vertices $G(E)$, we denote the
two spin states of the local system by $\pm$ and order the four states in Liouville space
as $++,--,+-,-+$. This means that states in Liouville space are vectors with 4 elements,
corresponding to operators in usual Hilbert space. Superoperators acting in Liouville space
are $4\times 4$-matrices. To parametrize an arbitrary $4\times 4$-matrix we decompose it
in four $2\times 2$-blocks, each of which can be decomposed in the basis of the unity matrix
$\sigma_0\equiv\mathbbm{1}_2$ and the three Pauli matrices $\sigma_i$ ($i=1,2,3\equiv x,y,z$). 
For an arbitrary $4\times 4$-matrix A, we introduce the following elegant tensor notation
\begin{equation}
\label{eq:matrix_general}
A\,=\,\sum_{i=0,1,2,3}\,\left(\begin{array}{cc}A_{11}^i\sigma_i & A_{12}^i\sigma_i \\ 
A_{21}^i\sigma_i & A_{22}^i\sigma_i \end{array}\right)
\,=\,\sum_{i=0,1,2,3}\,\left(\begin{array}{cc}A_{11}^i & A_{12}^i \\ 
A_{21}^i & A_{22}^i \end{array}\right)\otimes\sigma_i
\,=\,\sum_{i=0,1,2,3}\,A_i\otimes\sigma_i\quad,
\end{equation}
where $A_i$ is a $2\times 2$-matrix for all $i=0,1,2,3$. This notation has the advantage
that a product of two $4\times 4$-matrices $A$ and $B$ can be written as
\begin{equation}
\label{eq:matrix_product}
AB\,=\,\sum_{ij}\,(A_i\otimes\sigma_i)\,(B_j\otimes\sigma_j)
\,=\,\sum_{ij}\,(A_i B_j)\otimes(\sigma_i\sigma_j)
\,=\,(\underline{A}\cdot\underline{B})\otimes\mathbbm{1}_2\,+\,
i\,(\underline{A}\wedge\underline{B})\otimes\underline{\sigma}\quad,
\end{equation}
with $\underline{A}^T=(A_x,A_y,A_z)$, $\underline{B}^T=(B_x,B_y,B_z)$ and 
$\underline{\sigma}^T=(\sigma_x,\sigma_y,\sigma_z)$. The inverse of a matrix is
given by 
\begin{equation}
\label{eq:matrix_inversion}
A^{-1}\,=\,\sum_{i=0,1,2,3}\,A_i^{-1}\otimes\sigma_i\quad.
\end{equation}

Using this notation we get from the Hamiltonian the following matrix structure for
the bare Liouvillian and the bare vertices
\begin{align}
\label{eq:bare_Liouvillian}
L^{(0)}\,&=\,[H,\cdot]\,=\,\Delta\,
\left(\begin{array}{cc} 0 & 1 \\ 1 & 0 \end{array}\right)\otimes\tau_-\quad,\\
\label{eq:bare_vertex_average}
G^{(0)}\,&=\,\sum_p\,G^{pp,(0)}\,=\,{1\over 2}[\sigma_z,\cdot]\,=\,
\left(\begin{array}{cc} 0 & 0 \\ 0 & 1 \end{array}\right)\otimes\sigma_z\quad,\\
\label{eq:bare_vertex_tilde}
\tilde{G}^{(0)}\,&=\,\sum_p\,p\,G^{pp,(0)}\,=\,{1\over 2}\{\sigma_z,\cdot\}\,=\,
\left(\begin{array}{cc} 1 & 0 \\ 0 & 0 \end{array}\right)\otimes\sigma_z\quad,
\end{align}
where $p$ is the Keldysh index, $[\cdot,\cdot]$ denotes the commutator and
$\{\cdot,\cdot\}$ is the anti-commutator. Instead of $\mathbbm{1}_2$ and $\sigma_x$, we 
use the matrices $\tau_\pm$ defined by 
\begin{equation}
\label{eq:tau_matrices}
\tau_\pm\,=\,{1\over 2}(1\pm\sigma_x)\quad,\quad{\text{with}}\quad
\tau_\sigma\tau_{\sigma'}\,=\,\delta_{\sigma\sigma'}\,\tau_\sigma\quad,\quad
\tau_\sigma^{-1}\,=\,\tau_\sigma\quad,\quad
\sigma_z\tau_\sigma\,=\,\tau_{-\sigma}\sigma_z\quad.
\end{equation}
Using (\ref{eq:matrix_product}-\ref{eq:tau_matrices}), we get for the bare propagator
\begin{equation}
\label{eq:bare_propagator}
\Pi^{(0)}(E)\,=\,{1\over E-L^{(0)}}\,=\,\sum_\sigma\,\Pi^{(0),\sigma}\,\otimes\,\tau_\sigma \quad,\quad
\Pi^{(0),+}\,=\,{1\over E}\left(\begin{array}{cc} 1 & 0 \\ 0 & 1 \end{array}\right) \quad,\quad
\Pi^{(0),-}\,=\,{1\over 2}\sum_{\sigma=\pm}\,{1\over E-\sigma\Delta}
\left(\begin{array}{cc} 1 & \sigma \\ \sigma & 1 \end{array}\right)\quad,
\end{equation}
and for the sequence of two vertices
\begin{align}
\label{eq:product_G_G}
G^{(0)}\,\Pi^{(0)}(E+\omega)\,G^{(0)}\,&=\,
{1\over 2}\sum_{\sigma=\pm}\,{1\over E+\omega-\sigma\Delta}
\left(\begin{array}{cc} 0 & 0 \\ 0 & 1 \end{array}\right)\otimes\tau_+ \,+\,
{1\over E+\omega}\left(\begin{array}{cc} 0 & 0 \\ 0 & 1 \end{array}\right)\otimes\tau_- \quad,\\
\label{eq:product_G_tilde_G}
G^{(0)}\,\Pi^{(0)}(E+\omega)\,\tilde{G}^{(0)}\,&=\,
{1\over 2}\sum_{\sigma=\pm}\,{\sigma\over E+\omega-\sigma\Delta}
\left(\begin{array}{cc} 0 & 0 \\ 1 & 0 \end{array}\right)\otimes\tau_+ \quad.
\end{align}
The vertex $\tilde{G}^{(0)}$ can only occur if the symmetric part 
$\gamma^{p'}_s(\omega)=-\alpha p'\omega$ of some contraction connects this vertex with
a vertex $G^{(0)}$ standing left to it. If no other vertex stands between these two
vertices, we see from (\ref{eq:product_G_tilde_G}) that the propagator gives a contribution
$\sim{1\over \omega^2}$ for large $\omega$, i.e. together with the linear frequency 
dependence of $\gamma^{p'}_s(\omega)$ the frequency integration $\int d\omega$ involves an integrand
$\sim{1\over \omega}$ for large $\omega$. This gives the following contribution to the
effective Liouvillian for $E\equiv E+i0^+$ slightly above the real axis
\begin{align}
\nonumber
L^s\quad&=\quad 
\begin{picture}(10,10)
\put(5,-6){\includegraphics[height=0.8cm]{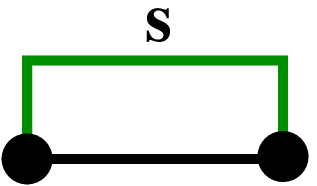}}
\end{picture}
\hspace{1.7cm}=\quad
-\alpha\,\int d\omega \,\omega\,G^{(0)}\,\Pi^{(0)}(E+\omega+i0^+)\,\tilde{G}^{(0)}\\
\nonumber
&=\quad
-{1\over 2}\alpha\,\sum_{\sigma=\pm}\,\int d\omega\,{\sigma\omega\over E+\omega-\sigma\Delta+i0^+}
\left(\begin{array}{cc} 0 & 0 \\ 1 & 0 \end{array}\right)\otimes\tau_+ \\
\nonumber
&=\quad
-{1\over 2}\alpha\,\sum_{\sigma=\pm}\,\int d\omega\,{-\sigma E+\Delta\over E+\omega-\sigma\Delta+i0^+}
\left(\begin{array}{cc} 0 & 0 \\ 1 & 0 \end{array}\right)\otimes\tau_+ \\
\nonumber
&=\quad
-{1\over 2}\alpha\,\sum_{\sigma=\pm}\,(-\sigma E+\Delta)(-i\pi)\,\int d\omega\, 
\delta(E+\omega-\sigma\Delta)\left(\begin{array}{cc} 0 & 0 \\ 1 & 0 \end{array}\right)\otimes\tau_+ \\
\label{eq:L_symmetric}
&=\quad
i\pi\,\alpha\,\Delta\,\left(\begin{array}{cc} 0 & 0 \\ 1 & 0 \end{array}\right)\otimes\tau_+ \quad.
\end{align}
We note that the frequency integral is not logarithmically divergent and can be directly calculated
for $D\rightarrow\infty$, whereas the sequence (\ref{eq:product_G_G}) together with the 
antisymmetric part of the contraction leads to a logarithmically divergent integral 
which has to be treated by RG. The term $L^s$ gives rise to a perturbative and energy independent 
contribution to the Liouvillian. It is not possible that the
symmetric contraction $\gamma^{p'}_s(\omega)$ crosses over more than one propagator, since otherwise 
the integrand will be $\sim {1\over\omega^2}$ for large $\omega$ and the integration contour can
be closed in the upper half and gives zero since all propagators and the symmetric contraction are 
analytic in the upper half. Furthermore, since $L^s L^{(0)}=L^s G^{(0)}=0$, the part $L^s$ can not
appear in any diagram involving more than two vertices. Therefore, in the universal limit 
$D\rightarrow\infty$, we can split the Liouvillian exactly as
\begin{equation}
\label{eq:Liouvillian_splitting}
L(E)\,=\,\hat{L}(E)\,+\,L^s \quad\text{with}\quad 
\hat{L}(E)\,=\,L_\Delta(E)\,+\,E\,L'(E)\quad,\quad
L^s\,=\,i\pi\,\alpha\,\Delta\,\left(\begin{array}{cc} 0 & 0 \\ 1 & 0 \end{array}\right)\otimes\tau_+ \quad,
\end{equation}
where $L_\Delta(E)$ and $L'(E)$ are logarithmic functions which can be determined from a diagrammatic
series involving only the vertex $G^{(0)}$ averaged over the Keldysh indices. Since these diagrams
involve always an even number of vertices, we find together with 
$\Pi^{(0)}(E)=\sum_\sigma\Pi^{(0)}_\sigma\otimes\tau_\sigma$, 
$G^{(0)}=\left(\begin{array}{cc} 0 & 0 \\ 0 & 1 \end{array}\right)\otimes\sigma_z$ and the 
algebra (\ref{eq:tau_matrices}), the form $L_\Delta(E)=\sum_\sigma L_\Delta^\sigma(E)\otimes\tau_\sigma$ and
$L'(E)=\sum_\sigma L'_\sigma(E)\otimes\tau_\sigma$, where each diagram can only contribute to the 
matrix elements $L_\Delta^\sigma(E)_{22}$ and $L'_\sigma(E)_{22}$. Since the bare quantities are given
by $L_\Delta^{(0)}=L^{(0)}$ and $L^{\prime,(0)}=0$, we can parametrize $L_\Delta(E)$ and $L'(E)$ in 
the form
\begin{align}
\label{eq:parametrization_L_Delta}
L_\Delta(E)\,&=\,\left(\begin{array}{cc} 0 & 0 \\ 0 & -i\Gamma^+_\Delta(E) \end{array}\right)\otimes\tau_+ 
\,+\,\left(\begin{array}{cc} 0 & \Delta \\ \Delta & -i\Gamma^-_\Delta(E) \end{array}\right)\otimes\tau_- \quad,\\ 
\label{eq:parametrization_L_prime}
L'(E)\,&=\,\sum_\sigma
\left(\begin{array}{cc} 0 & 0 \\ 0 & -i\Gamma_\sigma^\prime(E) \end{array}\right)\otimes\tau_\sigma \quad.
\end{align}
This gives for $\tilde{L}_\Delta(E)=Z'(E)L_\Delta(E)$ and $Z'(E)={1\over 1-L'(E)}$ the parametrization
\begin{align}
\label{eq:parametrization_L_tilde_Delta}
\tilde{L}_\Delta(E)\,&=\,\left(\begin{array}{cc} 0 & 0 \\ 0 & -i\Gamma_+(E) \end{array}\right)\otimes\tau_+ 
\,+\,\left(\begin{array}{cc} 0 & \Delta \\ Z_-(E)\Delta & -i\Gamma_-(E) \end{array}\right)\otimes\tau_- \quad,\\ 
\label{eq:parametrization_Z_prime}
Z'(E)\,&=\,\sum_\sigma
\left(\begin{array}{cc} 1 & 0 \\ 0 & Z_\sigma(E) \end{array}\right)\otimes\tau_\sigma \quad,
\end{align}
with $Z_\sigma(E)={1\over 1+i\Gamma_\sigma^\prime(E)}$ and $\Gamma_\sigma(E)=Z_\sigma(E)\Gamma_\Delta^\sigma(E)$.
Due to $L^s \hat{L}(E)=0$, we can write for the full propagator
\begin{equation}
\label{eq:full_propagator}
\Pi(E)\,=\,{1\over E-L(E)}\,=\,\hat{\Pi}(E)\,\left(1\,+\,L^s\,{1\over E}\right)\quad\text{with}\quad
\hat{\Pi}(E)\,=\,{1\over E-\hat{L}(E)}\,=\,\tilde{\Pi}(E)\,Z'(E) \quad,
\end{equation}
with $\tilde{\Pi}(E)={1\over E-\tilde{L}_\Delta(E)}$. In the RG equations, only the part $\hat{\Pi}(E)$ of
the propagator contributes.

In contrast to the Liouvillian, any diagram for the effective vertex $G(E)$ will involve 
an odd number of bare vertices, which leads to the general form 
$G(E)=\sum_{i=y,z}g_i(E)\left(\begin{array}{cc} 0 & 0 \\ 0 & 1 \end{array}\right)\otimes\sigma_i$ involving
the sector of the Pauli matrices $\sigma_{y,z}$. However, by inspecting the sum of mirrored diagrams, it
turns out that only the sector $\sigma_z$ remains. To see this consider a diagram of the form
\begin{equation}
\label{eq:g_diagram_1}
G^{(0)}\,\Pi^{(0),\sigma}_{X_1}\,G^{(0)}\,\Pi^{(0),-\sigma}_{X_2}\,\cdots\,
\Pi^{(0),\sigma}_{X_{2n-1}}\,G^{(0)}\,\Pi^{(0),-\sigma}_{X_{2n}}\,G^{(0)}\,=\,
A\,\otimes\,(\sigma_z\,\tau_\sigma\,\sigma_z\,\tau_{-\sigma})^n\,\sigma_z\,=\,
A\,\otimes\,(\tau_{-\sigma}\,\sigma_z)\quad,
\end{equation}
where $X_i$ is the set of frequency indices crossing over the $i$-th resolvent from
the left, and $\Pi^{(0),\sigma}_{X_i}=\Pi^{(0),\sigma}(E+\bar{\omega}_{X_i})$.
Adding the mirrored diagram
\begin{equation}
\label{eq:g_diagram_2}
G^{(0)}\,\Pi^{(0),-\sigma}_{X_{2n}}\,G^{(0)}\,\Pi^{(0),\sigma}_{X_{2n-1}}\,\cdots\,
\Pi^{(0),-\sigma}_{X_2}\,G^{(0)}\,\Pi^{(0),\sigma}_{X_1}\,G^{(0)}\,=\,
A\,\otimes\,(\sigma_z\,\tau_{-\sigma}\,\sigma_z\,\tau_{\sigma})^n\,\sigma_z\,=\,
A\,\otimes\,(\tau_{\sigma}\,\sigma_z)\quad,
\end{equation}
we get in total $A\otimes\sigma_z$. Therefore, the effective vertex is given by the parametrization
\begin{align}
\label{eq:parametrization_G}
G(E)\,=\,g(E)\,\left(\begin{array}{cc} 0 & 0 \\ 0 & 1 \end{array}\right)\otimes\sigma_z \quad.
\end{align}

From (\ref{eq:parametrization_L_tilde_Delta}) one can find the four eigenvalues $\lambda_k(E)$ of 
$\tilde{L}_\Delta(E)$ together with the projectors $P_k(E)$. A straightforward algebra gives the result
(we omit the energy argument $E$ in all expressions)
\begin{align}
\label{eq:eigenvalue_st}
\lambda_{\text{st}}\,=\,0 \quad\quad &, \quad\quad P_{\text{st}}\,Z'\,=\,  
\left(\begin{array}{cc} 1 & 0 \\ 0 & 0 \end{array}\right)\otimes\tau_+ \quad,\\
\label{eq:eigenvalue_0}
\lambda_0\,=\,-i\Gamma_+ \quad\quad &, \quad\quad P_0\,Z'\,=\,  
\left(\begin{array}{cc} 0 & 0 \\ 0 & Z_+ \end{array}\right)\otimes\tau_+ \quad,\\
\label{eq:eigenvalue_sigma}
\lambda_\sigma\,=\,-i{\Gamma_-\over 2}+\sigma\sqrt{Z_-\Delta^2-\Gamma_-^2/4} 
\quad\quad &, \quad\quad P_\sigma\,Z'\,=\,{\sigma\over 2\sqrt{Z_-\Delta^2-\Gamma_-^2/4}}\,
\left(\begin{array}{cc} \lambda_\sigma+i\Gamma_- & Z_-\Delta \\ 
Z_-\Delta & Z_-\lambda_\sigma \end{array}\right)\otimes\tau_- \quad.
\end{align}
With these quantities, the propagator $\hat{\Pi}(E+\omega)$ appearing in the RG equations  
can be expressed as
\begin{equation}
\label{eq:prop_rg}
\hat{\Pi}(E+\omega)\,\approx\,{1\over E+\omega-\tilde{L}_\Delta(E)}\,Z'(E)\,=\,
\sum_k\,{1\over E+\omega-\lambda_k(E)}\,P_k(E)\,Z'(E)\quad,
\end{equation}
where we have used the approximation (\ref{eq:prop_frequency}) neglecting the frequency dependence
in all logarithmic functions in leading order.

With the parametrization (\ref{eq:parametrization_G}-\ref{eq:prop_rg}) for the vertex
and the propagator, we now can easily evaluate the RG equations (\ref{eq:L_rg_firstorder_spinboson})
and (\ref{eq:G_rg_spinboson}) with $\Pi(E+\omega)\rightarrow\hat{\Pi}(E+\omega)$ and
the form (\ref{eq:res_contr_T=0}) for the derivatives of the antisymmetric part of the contraction 
at zero temperature. We get 
\begin{align}
\nonumber
{\partial \over \partial E}L_\Delta(E)
\,&=\,\int d\omega\,{d^2\gamma_a\over d\omega^2}(\omega)\,G(E)
{\omega-\tilde{L}_\Delta(E)\over E+\omega-\tilde{L}_\Delta(E)}\,Z'(E)\,G(E)
\,=\,-2\alpha\,G(E){\tilde{L}_\Delta(E)\over E-\tilde{L}_\Delta(E)}\,Z'(E)\,G(E)\\
\label{eq:L_Delta_rg_spinboson_evaluation}
\,&=\,-2\alpha\,\sum_k\,{\lambda_k(E)\over E-\lambda_k(E)}\,G(E)\,P_k(E)\,Z'(E)\,G(E)\quad,\\
\label{eq:L_prime_rg_spinboson_evaluation}
{\partial \over \partial E}L'(E)
\,&=\,\int d\omega\,{d^2\gamma_a\over d\omega^2}(\omega)\,G(E)
{1\over E+\omega-\tilde{L}_\Delta(E)}\,Z'(E)\,G(E)
\,=\,2\alpha\,\sum_k\,{1\over E-\lambda_k(E)}\,G(E)\,P_k(E)\,Z'(E)\,G(E)\quad,\\
\nonumber
{\partial \over \partial E}G(E)
\,&=\,-\int d\omega\,{d\gamma_a\over d\omega}(\omega)\,G(E)
{1\over E+\omega-\tilde{L}_\Delta(E)}\,Z'(E)\,G(E)\,{1\over E+\omega-\tilde{L}_\Delta(E)}\,Z'(E)\,G(E)\\
\label{eq:G_rg_spinboson_evaluation}
\,&=\,-\alpha\,\sum_{kk'}\,G(E)\,P_k(E)\,Z'(E)\,G(E)\,P_{k'}(E)\,Z'(E)\,G(E)
\,\int d\omega\,\text{sign}(\omega)\,{1\over E+\omega-\lambda_k(E)}\,{1\over E+\omega-\lambda_{k'}(E)}\quad,
\end{align}
and 
\begin{align}
\nonumber
{\partial \over \partial E}\tilde{L}_\Delta(E)
\,&=\,{\partial Z'\over \partial E}(E)\,L_\Delta(E)\,+\,Z'(E)\,{\partial L_\Delta\over \partial E}(E)\\
\label{eq:tilde_L_Delta_rg_spinboson_evaluation}
\,&=\,2\alpha\,\sum_k\,Z'(E)\,G(E)\,P_k(E)\,Z'(E)\,G(E)\,
{\tilde{L}_\Delta(E)-\lambda_k(E)\over E-\lambda_k(E)}\quad,\\
\label{eq:Z_prime_rg_spinboson_evaluation}
{\partial \over \partial E}Z'(E)
\,&=\,Z'(E)\,{\partial L'\over \partial E}\,Z'(E)
\,=\,2\alpha\,\sum_k\,Z'(E)\,G(E)\,P_k(E)\,Z'(E)\,G(E)\,Z'(E)\,{1\over E-\lambda_k(E)}\quad.
\end{align}
Using (we omit the argument $E$ everywhere)
\begin{align}
\label{eq:algebra_identities_1}
Z'GP_{\text{st}}Z'G\,&=\,0 \quad,\quad
Z'GP_0Z'G\,=\,Z_+ Z_- g^2 \left(\begin{array}{cc} 0 & 0 \\ 0 & 1 \end{array}\right)\otimes\tau_- \quad,\\
\label{eq:algebra_identities_2}
Z'GP_\sigma Z'G\,&=\,Z_+ Z_- g^2 {\sigma\lambda_\sigma\over 2\sqrt{Z_-\Delta^2-\Gamma_-^2/4}}\,
\left(\begin{array}{cc} 0 & 0 \\ 0 & 1 \end{array}\right)\otimes\tau_+ \quad,\\
\label{eq:algebra_identities_3}
GP_k Z'G P_{k'}Z'G\,&=\, Z_+ Z_- g^3 {\sigma\lambda_\sigma\over 2\sqrt{Z_-\Delta^2-\Gamma_-^2/4}}\,
\left(\begin{array}{cc} 0 & 0 \\ 0 & 1 \end{array}\right)\otimes
(\delta_{k\sigma}\delta_{k'0}\tau_+ + \delta_{k0}\delta_{k'\sigma}\tau_-)\sigma_z \quad,\\
\label{eq:algebra_identities_4}
&\hspace{-3cm}
\int d\omega\,\text{sign}(\omega)\,
{1\over E+\omega-\lambda_k}\,{1\over E+\omega-\lambda_{k'}}\,=\,
-2\,{1\over \lambda_k-\lambda_{k'}}\,\ln{E-\lambda_k\over E-\lambda_{k'}}\quad,
\end{align}
we obtain the RG equations
\begin{align}
\label{eq:s_rg_gamma_+-}
{\partial \Gamma_+\over \partial E}
\,&=\,i\alpha\,\tilde{g}\,\sum_\sigma\,{\sigma\lambda_\sigma\over\sqrt{\tilde{\Delta}^2-\Gamma_-^2/4}}\,
{\lambda_0-\lambda_\sigma\over E-\lambda_\sigma} \hspace{1.3cm}\quad,\quad
{\partial \Gamma_-\over \partial E}
\,=\,2\alpha\,\tilde{g}\,{\Gamma_--\Gamma_+\over E-\lambda_0}\quad,\\
\label{eq:s_rg_Z_+-}
{\partial Z_+\over \partial E}
\,&=\,\alpha\,Z_+\,\tilde{g}\,\sum_\sigma\,{\sigma\lambda_\sigma\over\sqrt{\tilde{\Delta}^2-\Gamma_-^2/4}}\,
{1\over E-\lambda_\sigma}\hspace{1cm}\quad,\quad
{\partial Z_-\over \partial E}
\,=\,2\alpha\,Z_-\,\tilde{g}\,{1\over E-\lambda_0} \quad,\\
\label{eq:s_rg_g}
{\partial g\over \partial E}
\,&=\,\alpha\,g\,\tilde{g}\,\sum_\sigma\,{\sigma\lambda_\sigma\over\sqrt{\tilde{\Delta}^2-\Gamma_-^2/4}}\,
{1\over \lambda_0-\lambda_\sigma}\ln{E-\lambda_0\over E-\lambda_\sigma}\quad,&&
\end{align}
where we have defined $\tilde{g}=Z_+ Z_- g^2$, and $\tilde{\Delta}=\sqrt{Z_-}\Delta$ denotes the
renormalized tunneling. The RG equations simplify considerably since we can
use $\tilde{\Delta}\gg \Gamma_-$, which is fulfilled during the whole RG flow since
$\Gamma_-\lesssim \alpha^2\tilde{\Delta}$ (see below).
Using $\lambda_\sigma\approx \sigma\tilde{\Delta}$ and 
$\sqrt{\tilde{\Delta}^2-\Gamma_-^2/4}\approx\tilde{\Delta}$, we get the final equations
\begin{align}
\label{eq:s_ap_rg_gamma_+-}
{\partial \Gamma_+\over \partial E}
\,&=\,i\alpha\,\tilde{g}\,\sum_\sigma\,{\lambda_0 - \lambda_\sigma \over E-\lambda_\sigma}
\quad\quad\hspace{0.5cm},\quad\quad
{\partial \Gamma_-\over \partial E}
\,=\,2\alpha\,\tilde{g}\,{\Gamma_--\Gamma_+\over E-\lambda_0}\quad,\\
\label{eq:s_ap_rg_Z_+-}
{\partial Z_+\over \partial E}
\,&=\,\alpha\,Z_+\,\tilde{g}\,\sum_\sigma\,{1\over E-\lambda_\sigma} 
\quad\quad\hspace{0.3cm},\quad\quad
{\partial Z_-\over \partial E}
\,=\,2\alpha\,Z_-\,\tilde{g}\,{1\over E-\lambda_0}\quad,\\
\label{eq:s_ap_rg_g}
{\partial g\over \partial E}
\,&=\,\alpha\,g\,\tilde{g}\,\sum_\sigma\,{1\over \lambda_0-\lambda_\sigma}\,
\ln{E-\lambda_0 \over E-\lambda_\sigma}\quad,\\
\label{eq:s_ap_rg_g_tilde}
{\partial \tilde{g}\over \partial E}
\,&=\,\alpha\,\tilde{g}^2\,\sum_\sigma\,\left({1\over E-\lambda_\sigma}\,+\,{1\over E-\lambda_0}
\,+\,{2\over \lambda_0-\lambda_\sigma}\,
\ln{E-\lambda_0 \over E-\lambda_\sigma}\right)\quad,
\end{align}
together with the initial conditions at $E=iD$
\begin{equation}
\label{eq:initial}
\Gamma_\pm|_{E=iD}\,=\,0\quad,\quad
Z_\pm|_{E=iD}\,=\,1\quad,\quad
g|_{E=iD}\,=\,\tilde{g}|_{E=iD}\,=\,1\quad.\quad
\end{equation}

To solve the RG equations we first consider the regime of {\it large energies} 
$|E|\gg |\lambda_0|,|\lambda_\pm|$. In this regime we get
\begin{align}
\label{eq:large_rg_gamma_+-}
{\partial \Gamma_+\over \partial E}
\,&\approx\,2\alpha\,\tilde{g}\,\left({\Gamma_+\over E}\,-\,i{Z_-\Delta^2\over E^2}\right)
\hspace{1cm},\quad
{\partial \Gamma_-\over \partial E}
\,\approx\,2\alpha\,\tilde{g}\,{\Gamma_--\Gamma_+\over E}\quad,\\
\label{eq:large_rg_Z_g}
{\partial Z_\pm\over \partial E}
\,&\approx\,2\alpha\,Z_\pm\,\tilde{g}\,{1\over E} 
\hspace{3cm},\quad
{\partial \tilde{g}\over \partial E}
\,\approx\,{2\over 3}\,\alpha\,\tilde{g}^2\,\sum_\sigma\,{(\lambda_0-\lambda_\sigma)^2 \over E^3}\quad,
\end{align}
which leads to the result
\begin{align}
\label{eq:large_energies_Gamma}
\Gamma_+\,&\approx\,2i\alpha\,{Z_-\Delta^2\over E}\quad,\quad
\Gamma_-\,\approx\,4i\alpha^2\,{Z_-\Delta^2\over E}\,\approx\,2\alpha\,\Gamma_+\quad,\quad\\
\label{eq:large_energies_Z_g}
Z_\pm\,&\approx\,\left({-iE\over D}\right)^{2\alpha}\quad,\quad
\tilde{g}\,\approx\,1\,-\,{2\over 3}\,\alpha\, {Z_-\Delta^2\over E^2}\quad.
\end{align}
Obviously, we have resummed in this solution all powers of logarithmic divergencies 
$\sim(\alpha\ln(D/E))^n$. This defines the {\it poor man scaling regime}, which, in time space, corresponds
to the {\it short-time regime}. Defining a low energy scale by $\Omega=\Delta(\Omega/D)^\alpha$, 
we can write the solution for $Z_\pm$ as
\begin{equation}
\label{eq:Z_large_energies}
Z_\pm\,\approx\,\left({\Omega\over\Delta}\right)^2\,\left({-iE\over\Omega}\right)^{2\alpha}
\quad\text{with}\quad
\Omega\,=\,\Delta\,\left({\Omega\over D}\right)^\alpha\,=\,
\Delta\,\left({\Delta\over D}\right)^{\alpha\over 1-\alpha}\quad,
\end{equation}
such that we obtain a universal function in terms of the effective tunneling $\Omega$. Since
$\Gamma_-\lesssim \alpha^2\tilde{\Delta}$, we disregard it in the following.

We next consider the regime of {\it intermediate energies}, 
where $\alpha\ln{|\lambda_k|\over|E-\lambda_k|}\ll 1$ ($k=0,\pm$).
In this regime we can solve the RG equations perturbatively in $\alpha$ with the result
\begin{align}
\label{eq:intermediate_energies_Gamma_+_g}
\Gamma_+\,&\approx\,i\alpha\,\sum_\sigma\,(\lambda_0-\lambda_\sigma)\,
\ln{-i(E-\lambda_\sigma)\over\Omega}\quad,\quad
\tilde{g}\,=\,1\,+\,O(\alpha)\quad,\quad
g\,=\,\left({\Delta\over\Omega}\right)^2\,+\,O(\alpha)\quad,\\
\label{eq:intermediate_energies_Z}
Z_+\,&\approx\,\left({\Omega\over \Delta}\right)^2\,\left(1\,+\,
\alpha\,\sum_\sigma\,\ln{-i(E-\lambda_\sigma)\over\Omega}\right)\quad,\quad
Z_-\,\approx\,\left({\Omega\over \Delta}\right)^2\,\left(1\,+\,
2\alpha\,\ln{-i(E-\lambda_0)\over\Omega}\right)\quad,
\end{align}
where all integration constants have been fixed by comparison with the solution at
high energies. In this solution we have resummed all powers of logarithmic divergencies
$\sim (\alpha\ln(D/\Omega))^n$, and have expanded in the other small logarithmic functions.
It defines the {\it weak-coupling expansion regime}, which, in time space, corresponds to the 
{\it intermediate time regime}. This weak-coupling expansion is equivalent to the one developed
in Ref.~[17]. 

From the perturbative solution at intermediate energies, we can already derive the
leading order result for the real and imaginary parts of the pole positions, defined by
$z_k=\lambda_k(z_k)$ (note that $z_k=z_k^p$ since
there is no chemical potential in the present problem), with $\lambda_0=-i\Gamma_+$ and
$\lambda_\sigma=\sigma\sqrt{Z_-}\Delta$. For intermediate energies, we 
can approximately set $\lambda_k(E)\approx z_k(1+O(\alpha))$ in 
(\ref{eq:intermediate_energies_Gamma_+_g}) and (\ref{eq:intermediate_energies_Z}), and find directly
$z_0=-i\Gamma_+(z_0)\sim O(\alpha)$ and 
$z_\sigma=\sigma\sqrt{Z_-(z_\sigma)}\Delta=\sigma\Omega(1+O(\alpha))$. Using this result in 
(\ref{eq:intermediate_energies_Gamma_+_g}) and (\ref{eq:intermediate_energies_Z}) to calculate
the $O(\alpha)$ correction, we find
\begin{align}
\label{eq:z_0_determination}
iz_0&=\Gamma_+(z_0)\approx i\alpha\sum_\sigma (z_0-z_\sigma) \ln{-i(z_0-z_\sigma)\over\Omega}
\approx -i\alpha \Omega \sum_\sigma \sigma\ln(i\sigma)=\pi\alpha\Omega\quad,\\
\label{eq:z_+_determination}
{z_+\over\Delta}&=\sqrt{Z_-(z_+)}\approx {\Omega\over\Delta}\left(1+\alpha\ln{-i(z_+-z_0)\over\Omega}\right)
\approx{\Omega\over\Delta}\left(1+\alpha\ln(-i)\right)={\Omega\over\Delta}(1-i{\pi\over 2}\alpha)\quad,
\end{align}
i.e. with $z_-=-z_+^*$, the pole positions are given by
\begin{equation}
\label{eq:poles}
z_0\,=\,-i\Gamma\quad,\quad
z_\pm\,=\,\pm\Omega\,-\,i{\Gamma\over 2}\quad{\text{with}}\quad
\Gamma\,=\,\pi\alpha\Omega\quad.
\end{equation}

Finally, we consider the regime of {\it small energies}, where the Laplace variable $E$ approaches
one of the singularities $z_k$, such that $\alpha\ln{|E-z_k|\over\Omega}\sim O(1)$. First, we note
that the RG equation (\ref{eq:s_ap_rg_g}) for the vertex function $g(E)$ leads to a very weak
logarithmic correction $\sim\alpha (E-z_k)\ln(E-z_k)$ close to the singularities which can be neglected. 
Therefore, we take the constant value
$g\approx (\Delta/\Omega)^2$ at intermediate {\it and} small energies. In contrast, the vertex 
function $\tilde{g}=Z_+ Z_- g^2$ behaves very differently. It is approximately a constant $\tilde{g}\approx 1$ 
for high and intermediate energies, but has strong logarithmic corrections $\sim\alpha\ln(E-z_k)$
close to the singularities, which arise from corresponding singularities of the Z-factors $Z_\pm$.
We start with the fixed point analysis around $z_0$, where $\Gamma_+(E)\approx \Gamma$ and
$Z_+(E)\approx (\Omega/\Delta)^2$ behave smoothly. In contrast, $Z_-(E)$ has a logarithmic singularity,
which can be determined from the RG equation (\ref{eq:s_ap_rg_Z_+-})
\begin{align}
\label{eq:fixed_point_analysis_Z_-}
{\partial Z_-\over \partial E}
&\approx 2\alpha Z_- \tilde{g} {1\over E-z_0}
= 2\alpha Z_-^2 Z_+ g^2 {1\over E-z_0}
\approx 2\alpha Z_-^2 \left({\Delta\over\Omega}\right)^2 {1\over E-z_0}\quad\\
&\Rightarrow\quad
{1\over Z_-(E)}\,\approx\,\text{const}\,-\,2\alpha\,
\left({\Delta\over\Omega}\right)^2 \ln{-i(E-z_0)\over\Omega}  \quad.
\end{align}
Fixing the integration constant by comparing with the solution (\ref{eq:intermediate_energies_Z}) at
intermediate energies, we find for $E$ close to $z_0$
\begin{equation}
\label{eq:small_energies_z_0}
g(E)\,\approx\,\left({\Delta\over\Omega}\right)^2 \quad,\quad
\Gamma_+(E)\,\approx\,\Gamma \quad,\quad
Z_+(E)\,\approx\,\left({\Omega\over\Delta}\right)^2 \quad,\quad
Z_-(E)\,\approx\,\left({\Omega\over\Delta}\right)^2\,{1\over 1-2\alpha\ln{-i(E-z_0)\over\Omega}} \quad.
\end{equation}
Close to the singularity $z_+$, we get $Z_-(E)\approx (\Omega/\Delta)^2$ but $\Gamma_+(E)$ and
$Z_+(E)$ have strong logarithmic corrections, which can be determined from the RG equations
(\ref{eq:s_ap_rg_gamma_+-}) and (\ref{eq:s_ap_rg_Z_+-}) in the following way
\begin{align}
\nonumber
{\partial Z_+\over \partial E}
&\approx \alpha Z_+ \tilde{g} {1\over E-z_+}
= \alpha Z_+^2 Z_- g^2 {1\over E-z_+}
\approx \alpha Z_+^2 \left({\Delta\over\Omega}\right)^2 {1\over E-z_+}\quad\\
\label{eq:fixed_point_analysis_Z_+}
&\Rightarrow\quad
{1\over Z_+(E)}\,\approx\,\text{const}\,-\,\alpha\,
\left({\Delta\over\Omega}\right)^2 \ln{-i(E-z_+)\over\Omega}  
\quad\overset{(\ref{eq:intermediate_energies_Z})}{\Rightarrow}\quad
Z_+(E)\,\approx\,\left({\Omega\over\Delta}\right)^2\,{1\over 1-\alpha\ln{-i(E-z_+)\over\Omega}} \quad,\\
\nonumber
{\partial \Gamma_+\over \partial E}
&\approx i\alpha \tilde{g} {\lambda_0-z_+\over E-z_+}
\approx -i\alpha Z_+ Z_- g^2 {\Omega\over E-z_+}
\approx -i\alpha Z_+\left({\Delta\over\Omega}\right)^2 {\Omega\over E-z_+}
\overset{(\ref{eq:fixed_point_analysis_Z_+})}{\approx} 
-i\alpha {1\over 1-\alpha\ln{-i(E-z_+)\over\Omega}}{\Omega\over E-z_+}\quad\\
\label{eq:fixed_point_analysis_Gamma_+}
&\Rightarrow\quad
\Gamma_+(E)\,\approx\,\text{const}\,+\,i\Omega\,
\ln\left(1-\alpha\ln{-i(E-z_+)\over\Omega}\right)
\quad\overset{(\ref{eq:intermediate_energies_Gamma_+_g})}{\Rightarrow}\quad
\Gamma_+(E)\,\approx\,{\Gamma\over 2}\,+\,i\Omega\,
\ln\left(1-\alpha\ln{-i(E-z_+)\over\Omega}\right) \quad,
\end{align}
where again the integration constants have been fixed by comparison with the solutions at
intermediate energies. Using a similar analysis close to $z_-$, we obtain for $E$ close
to $z_\sigma$ the result
\begin{align}
\label{eq:small_energies_z_sigma_g_Gamma_+}
g(E)\,&\approx\,\left({\Delta\over\Omega}\right)^2 \quad,&\quad
\Gamma_+(E)\,&\approx\,{\Gamma\over 2}\,+\,i\,\sigma\,\Omega\,
\ln\left(1\,-\,\alpha\,\ln{-i(E-z_\sigma)\over\Omega}\right) \quad,\\
\label{eq:small_energies_z_sigma_Z}
Z_+(E)\,&\approx\,\left({\Omega\over\Delta}\right)^2\,{1\over 1\,-\,\alpha\,\ln{-i(E-z_\sigma)\over\Omega}} 
\quad,&\quad
Z_-(E)\,&\approx\,\left({\Omega\over\Delta}\right)^2 \quad.
\end{align}

With these results we can also evaluate the RG equation (\ref{eq:delta_L_rg_spinboson})
for the jump $\delta L(z_k-ix)$ of the Liouvillian, which we parametrize as 
\begin{equation}
\label{eq:delta_L_parametrization}
\delta L(E)\,=\,-i\,\sum_\sigma\,\delta\gamma_\sigma(E)\,
\left(\begin{array}{cc} 0 & 0 \\ 0 & 1 \end{array}\right)\otimes\tau_\sigma\quad.
\end{equation}
Using the algebra (\ref{eq:algebra_identities_1})
and (\ref{eq:algebra_identities_2}) together with our results 
(\ref{eq:intermediate_energies_Gamma_+_g}), (\ref{eq:intermediate_energies_Z}),
(\ref{eq:small_energies_z_0}), (\ref{eq:small_energies_z_sigma_g_Gamma_+}) and 
(\ref{eq:small_energies_z_sigma_Z}) at intermediate and small energies, we get 
for $x\lesssim \Omega$ from (\ref{eq:delta_L_rg_spinboson})
\begin{align}
\label{eq:delta_L_evaluation_+}
-i{\partial\over\partial x}\delta\gamma_+(z_\sigma-ix)&=
-4\pi\alpha\theta(x)Z_-(z_\sigma-ix) g(z_\sigma-ix)^2{1\over 2}
\approx -2\pi\alpha\theta(x)\left({\Delta\over\Omega}\right)^2\quad,\\ 
\label{eq:delta_L_evaluation_-}
-i{\partial\over\partial x}\delta\gamma_-(z_0-ix)&=
-4\pi\alpha\theta(x)Z_+(z_0-ix) g(z_0-ix)^2
\approx -4\pi\alpha\theta(x)\left({\Delta\over\Omega}\right)^2\quad,
\end{align}
with the solution
\begin{equation}
\label{eq:delta_L_result}
\delta\gamma_+(z_\sigma-ix)\,\approx\,
-2\pi i\,\alpha\,\left({\Delta\over\Omega}\right)^2\,x\,\theta(x)\quad,\quad
\delta\gamma_-(z_0-ix)\,\approx\, 
-4\pi i\,\alpha\,\left({\Delta\over\Omega}\right)^2\,x\,\theta(x)\quad.
\end{equation}
 
{\bf Time evolution.}--- With the results for the Liouvillian we now can evaluate the
time evolution. We start with the {\it short time regime} $\Omega t\ll 1$. Using (\ref{eq:short_times}),
(\ref{eq:parametrization_Z_prime}) and (\ref{eq:Z_large_energies}), we find
\begin{equation}
\label{eq:short_times_spinboson}
\rho_t\,\approx\,\left(\begin{array}{cc} 1 & 0 \\ 
0 & \left({1\over Dt}\right)^{2\alpha} \end{array}\right)\otimes\mathbbm{1}_2\,\rho_{t=0}\quad.
\end{equation}

The {\it intermediate and long time regime} $\Omega t\gtrsim 1$ is based on the pole and branch cut contributions given by 
(\ref{eq:branching_pole}) and (\ref{eq:branching_point}), respectively. Thereby, 
we have to consider that the full propagator (\ref{eq:full_propagator})
involves the correction $\hat{\Pi}(E)L^s{1\over E}$. Since $L^s\sim O(\alpha)$, this leads
to a negligible $O(\alpha^2)$-correction to the branch cut contributions (\ref{eq:branching_point})
but the pole contribution (\ref{eq:branching_pole}) changes to
\begin{align}
\label{eq:time_pole_zero_general}
\rho_t^{\text{st},p}\,&=\,\rho_{\text{st}}\,=\,
\left(P_{\text{st}}(0)\,Z'(0)\,+\,\hat{\Pi}(0)\,L^s\right)\,\rho_{t=0}\,\approx\,
\left(P_{\text{st}}(0)\,Z'(0)\,-\,\sum_{k=0,\pm}\,P_k(0)\,Z'(0)\,L^s\,{1\over z_k}\right)\,\rho_{t=0}\quad,\\
\label{eq:time_pole_unzero_general}
\rho_t^{k,p}\,&=\,e^{-iz_k t}\,P_k(z_k)\,Z'(z_k)\,\left(1\,+\,L^s\,{1\over z_k}\right)\,\rho_{t=0}
\quad \text{for}\quad k=0,\pm\quad,
\end{align}
where we have used $P_{\text{st}}Z'L^s=0$, $\lambda_k(0)\approx z_k$ and the fact 
that all poles are isolated up to leading order truncation.
Using the form (\ref{eq:Liouvillian_splitting}) for $L^s$ together with the results
(\ref{eq:eigenvalue_st}-\ref{eq:eigenvalue_sigma}) for the projectors (where we neglect
$\Gamma_-$), we find
\begin{equation}
\label{eq:projector_Ls}
P_{\text{st}}Z'L^s\,=\,0\quad,\quad
P_0 Z' L^s\,=\,i\pi\alpha\Delta Z_+
\left(\begin{array}{cc} 0 & 0 \\ 1 & 0 \end{array}\right)\otimes\tau_+ \quad,\quad
P_\sigma Z' L^s\,=\,0\quad.
\end{equation}
Inserting these results in (\ref{eq:time_pole_zero_general}) and (\ref{eq:time_pole_unzero_general}), 
we find together with $Z_+(0)\approx Z_+(z_0)\approx Z_-(z_\sigma)\approx (\Omega/\Delta)^2$ 
and (\ref{eq:poles})
\begin{align}
\label{eq:time_pole_st}
\rho_t^{\text{st},p}&\,=\,\rho_{\text{st}}
\approx\left\{
\left(\begin{array}{cc} 1 & 0 \\ 0 & 0 \end{array}\right)-{i\pi\alpha\Delta Z_+(0)\over z_0}
\left(\begin{array}{cc} 0 & 0 \\ 1 & 0 \end{array}\right)
\right\}\otimes\tau_+\rho_{t=0}
\approx
\left(\begin{array}{cc} 1 & 0 \\ \Omega/\Delta & 0 \end{array}\right)\otimes\tau_+\rho_{t=0}
\,=\,{1\over 2}\left(\begin{array}{c} 1 \\ 1 \\ \Omega/\Delta \\ \Omega/\Delta \end{array}\right)
\quad,\\
\label{eq:time_pole_zero}
\rho_t^{0,p}&\,=\,e^{-iz_0t}Z_+(z_0)\left\{
\left(\begin{array}{cc} 0 & 0 \\ 0 & 1 \end{array}\right)+{i\pi\alpha\Delta \over z_0}
\left(\begin{array}{cc} 0 & 0 \\ 1 & 0 \end{array}\right)
\right\}\otimes\tau_+\rho_{t=0}
\,\approx\,
e^{-iz_0t}\left(\begin{array}{cc} 0 & 0 \\ -\Omega/\Delta & (\Omega/\Delta)^2 \end{array}\right)
\otimes\tau_+\rho_{t=0}
\quad,\\
\nonumber
\rho_t^{\sigma,p}&=e^{-iz_\sigma t}{\sigma \over 2\sqrt{Z_-(z_\sigma)}\Delta}
\left(\begin{array}{cc} \sigma\sqrt{Z_-(z_\sigma)}\Delta & Z_-(z_\sigma)\Delta \\ 
Z_-(z_\sigma)\Delta & \sigma Z_-(z_\sigma)\sqrt{Z_-(z_\sigma)}\Delta  \end{array}\right)
\otimes\tau_-\rho_{t=0}\\
\label{eq:time_pole_sigma}
&\approx\,
e^{-iz_\sigma t}{1\over 2}\left(\begin{array}{cc} 1 & \sigma\Omega/\Delta \\ 
\sigma\Omega/\Delta & (\Omega/\Delta)^2 \end{array}\right)
\otimes\tau_-\rho_{t=0}
\quad.
\end{align}

Finally, by using the algebra of the projectors $P_k Z'$ and the jump $\delta L$ according to
(\ref{eq:eigenvalue_st}-\ref{eq:eigenvalue_sigma}) and (\ref{eq:delta_L_parametrization}),
we can write the branch cut contribution (\ref{eq:branching_point}) as
\begin{align}
\label{eq:bc_evaluation_0}
\rho_t^{0,b}\,&=\,e^{-iz_0t}\,{1\over 2\pi}\,\sum_{\sigma\sigma'}\,\int_0^\infty dx\,
{e^{-xt}\over (z_0-ix-\bar{\lambda}^0_\sigma)\,(z_0-ix-\bar{\lambda}^0_{\sigma'})}\,
\bar{P}_\sigma^0\,\bar{Z}^{\prime 0}\,\delta L(z_0-ix)\,\bar{P}_{\sigma'}^0\,\bar{Z}^{\prime 0}
\,\rho_{t=0} \quad,\\
\label{eq:bc_evaluation_sigma}
\rho_t^{\sigma,b}\,&=\,e^{-iz_\sigma t}\,{1\over 2\pi}\,\sum_{\sigma\sigma'}\,\int_0^\infty dx\,
{e^{-xt}\over (z_\sigma-ix-\bar{\lambda}^\sigma_0)^2}\,
\bar{P}^\sigma_0\,\bar{Z}^{\prime \sigma}\,\delta L(z_\sigma-ix)\,\bar{P}^{\sigma'}_0\,\bar{Z}^{\prime \sigma}
\,\rho_{t=0} \quad.
\end{align}
For $\Omega t\gg 1$ we can use 
\begin{equation}
\label{eq:bc_resolvents}
z_0\,-ix\,-\,\bar{\lambda}^0_\sigma\,\approx\, -\bar{\lambda}^0_\sigma\,\approx\,
-\sigma\,\sqrt{\bar{Z}_-^0}\,\Delta \quad,\quad
z_\sigma\,-ix\,-\,\bar{\lambda}_0^\sigma\,\approx\,z_\sigma\,+\,i\,\bar{\Gamma}_+^\sigma \quad,
\end{equation}
where $\bar{Z}_-^0=\bar{Z}_-(z_0-i/t)$ and $\bar{\Gamma}_+^\sigma=\bar{\Gamma}_+(z_\sigma-i/t)$ can
be calculated from (\ref{eq:small_energies_z_0}) and (\ref{eq:small_energies_z_sigma_g_Gamma_+}) as
\begin{equation}
\label{eq:bc_Z-,Gamma+}
\bar{Z}_-^0\,\approx\,\left({\Omega\over\Delta}\right)^2\,{1\over 1+2\alpha\ln(\Omega t)} \quad,\quad
\bar{\Gamma}_+^\sigma\,\approx\,{\Gamma\over 2}\,+\,i\,\sigma\,\Omega\,
\ln\left(1\,+\,\alpha\,\ln(\Omega t)\right) \quad.
\end{equation}
Furthermore, due to the algebra of the projectors $P_k Z'$ and the jump $\delta L$, we can use
\begin{align}
\label{eq:bc_algebra_projectors_sigma}
P_\sigma\,Z'\,\delta L\,P_{\sigma'}\,Z'\,&=\,-{i\over 4}\,\delta\gamma_-\,Z_-\,
\left(\begin{array}{cc} \sigma\sigma' & \sigma\sqrt{Z_-} \\ 
\sigma'\sqrt{Z_-} & Z_- \end{array}\right)\otimes\tau_- \quad,\\
\label{eq:bc_algebra_projectors_0}
P_0\,Z'\,\delta L\,P_0\,Z'\,&=\,-i\,\delta\gamma_+\,(Z_+)^2\,
\left(\begin{array}{cc} 0 & 0 \\ 0 & 1 \end{array}\right)\otimes\tau_+ \quad.
\end{align}
Inserting (\ref{eq:bc_resolvents}-\ref{eq:bc_algebra_projectors_0}) in 
(\ref{eq:bc_evaluation_0}) and (\ref{eq:bc_evaluation_sigma}),
and using the result (\ref{eq:delta_L_result}) for $\delta\gamma_-(z_0-ix)$ and 
$\delta\gamma_+(z_\sigma-ix)$, we obtain
\begin{align}
\nonumber
\rho_t^{0,b}\,&\approx\,e^{-iz_0t}\,\left(-{i\over 2\pi\Delta^2}\right)\,
\int_0^\infty dx\,e^{-xt}\,\delta\gamma_-(z_0-ix)\,
\left(\begin{array}{cc} 1 & 0 \\ 0 & 0 \end{array}\right)\otimes\tau_-\,\rho_{t=0} \\
\nonumber
&=\,e^{-iz_0t}\,\left(-{2\alpha\over \Omega^2}\right)\,
\int_0^\infty dx\,e^{-xt}\,x\,
\left(\begin{array}{cc} 1 & 0 \\ 0 & 0 \end{array}\right)\otimes\tau_-\,\rho_{t=0} \\
\label{eq:bc_0}
&=\,e^{-iz_0t}\,\left(-{2\alpha\over (\Omega t)^2}\right)\,
\left(\begin{array}{cc} 1 & 0 \\ 0 & 0 \end{array}\right)\otimes\tau_-\,\rho_{t=0} \quad,\\
\nonumber
\rho_t^{\sigma,b}\,&\approx\,e^{-iz_\sigma t}\,\left(-{i\over 2\pi}\right)\,
\left({\bar{Z}_+^\sigma\over z_\sigma+i\Gamma_+^\sigma}\right)^2\,
\int_0^\infty dx\,e^{-xt}\,\delta\gamma_+(z_\sigma-ix)\,
\left(\begin{array}{cc} 0 & 0 \\ 0 & 1 \end{array}\right)\otimes\tau_+\,\rho_{t=0} \\
\nonumber
&=\,e^{-iz_\sigma t}\,\left(-{\alpha\,f_t\over \Delta^2}\right)\,
\int_0^\infty dx\,e^{-xt}\,x\,
\left(\begin{array}{cc} 0 & 0 \\ 0 & 1 \end{array}\right)\otimes\tau_+\,\rho_{t=0} \\
\label{eq:bc_sigma}
&=\,e^{-iz_\sigma t}\,\left(-{\alpha\,f_t\over (\Delta t)^2}\right)\,
\left(\begin{array}{cc} 0 & 0 \\ 0 & 1 \end{array}\right)\otimes\tau_+\,\rho_{t=0} \quad,
\end{align}
where we have defined the logarithmic scaling function
\begin{equation}
\label{eq:sm_f_t}
f_t\,=\,\left({1\over (1+\alpha\ln(\Omega t))(1-\ln(1+\alpha\ln(\Omega t)))}\right)^2 \quad.
\end{equation}

Eqs.~(\ref{eq:time_pole_st}-\ref{eq:time_pole_sigma}) and (\ref{eq:bc_0}-\ref{eq:sm_f_t}) are
the final results for the time evolution in the regime $\Omega t\gg 1$. We note that the 
branch cut integrals (\ref{eq:bc_evaluation_0}) and (\ref{eq:bc_evaluation_sigma}) can
also be calculated exactly in terms of exponential integrals, extending the 
applicability range to the time regime $\Omega t \gtrsim 1$.


\begin{thebibliography}{99}
\bibitem{comment_a}
At finite temperature $T$ it turns out that the branch cuts are replaced by 
an infinite series of poles separated by $2\pi T$.
\bibitem{comment_0}
We note that the pole $z^p_{\text{st}}=0$ for $k=\text{st}$ is usually not a 
branching pole but an isolated pole and fullfils 
$P_{\text{st}}(E)Z'(E)=P_{\text{st}}(E)=|x_{\text{st}}(E)\rangle\text{Tr}$.\cite{SM}
\bibitem{comment_1}
For special resonant conditions, the time regime $1/|z_n|\ll t\ll 1/|z_n-z_{k,k'}^p|$
is possible, where $x$ dominates in the resolvents and different scaling can be
obtained for intermediate times, see Ref.~\cite{Pletyukhov_Schuricht_Schoeller_10}
for the nonequilibrium Kondo model.
\bibitem{comment_2}
We note that the exponential part of the branch cut contributions can only 
be obtained from self-consistent Born approximation. Within RTRG, it can be 
shown generically that exponential decay is expected in all orders of 
perturbation theory if the symmetric part of the reservoir contraction is an
analytic function.\cite{Pletyukhov_Schuricht_Schoeller_10}
\end{thebibliography}
\end{document}